\documentclass[aps,prx,twocolumn,superscriptaddress,showpacs,floatfix]{revtex4-1}
\usepackage{amsmath,amssymb,graphicx}

\usepackage[utf8]{inputenc}
\usepackage[T1]{fontenc}
\usepackage{xcolor}

\IfFileExists{newtxtext.sty}
   {\usepackage{newtxtext,newtxmath}}
   {\IfFileExists{stix.sty}
      {\usepackage{stix}}
      {\IfFileExists{mathptmx.sty}
      {\usepackage{mathptmx}}{} } }

\usepackage{textcomp}

\usepackage{bm}

\IfFileExists{siunitx.sty}{\usepackage{booktabs,siunitx}}{}

\pdfoutput=1
\usepackage{color}
\definecolor{LinkColor}{rgb}{0.256,0.439,0.588}
\usepackage{hyperref}
\hypersetup{
   pdfauthor={Xiao Yan Xu, Kai Sun, and Zi Yang Meng},
   pdftitle={Non-Fermi-liquid at (2+1)d ferromagnetic quantum critical point},
   colorlinks=true,
   citecolor=LinkColor,
   linkcolor=LinkColor,
   urlcolor=LinkColor
}

\renewcommand{\vec}[1]{\mathbf{#1}}

\usepackage{pifont}

\begin{document}

\title{Non-Fermi-liquid at (2+1)d ferromagnetic quantum critical point}

\author{Xiao Yan Xu}
\affiliation{Beijing National Laboratory for Condensed Matter Physics and Institute
of Physics, Chinese Academy of Sciences, Beijing 100190, China}
\author{Kai Sun}
\affiliation{Department of Physics, University of Michigan, Ann Arbor, MI 48109, USA}
\author{Yoni Schattner}
\affiliation{Department of Condensed Matter Physics, Weizmann Institute of Science, Rehovot, Israel 76100}
\author{Erez Berg}
\affiliation{Department of Condensed Matter Physics, Weizmann Institute of Science, Rehovot, Israel 76100}
\author{Zi Yang Meng}
\affiliation{Beijing National Laboratory for Condensed Matter Physics and Institute
of Physics, Chinese Academy of Sciences, Beijing 100190, China}

\begin{abstract}
%EBWe develop a new quantum Monte Carlo setup to investigate itinerant quantum critical points (QCPs)
%in an unbiased manner.
We construct a two-dimensional lattice model of fermions coupled to Ising ferromagnetic critical fluctuations. Using extensive sign-problem-free
quantum Monte Carlo simulations, we show that the model realizes a continuous itinerant quantum phase transition.
In comparison with other similar itinerant quantum critical points (QCPs), our QCP shows much weaker superconductivity tendency with no superconducting state down to the lowest temperature investigated, hence making the system a good platform for the exploration of quantum critical fluctuations.
%study a two-dimensional lattice model that exhibits an itinerant ferromagnetic quantum critical point (QCP) using sign problem-free quantum Monte Carlo simulations.
%The model consists of fermions that form a Fermi surface, coupled to ferromagnetic Ising fluctuations.
%EB By constructing a ferromagnetic QCP with Fermi surface coupled
%to ferromagnetic Ising critical fluctuations, we have successfully realized a continuous quantum phase transition
%of itinerant ferromagnetism in (2+1)d.
Remarkably, clear signatures of non-Fermi-liquid behavior in the fermion propagators are observed at the QCP. %quantum critical point (QCP).
The critical  fluctuations at the QCP partially resemble Hertz-Millis-Moriya behavior. However,
careful scaling analysis reveals that the QCP belongs to a different universality class,
deviating from both (2+1)d Ising and Hertz-Millis-Moriya predictions.
\end{abstract}

\date{\today}

\maketitle

\section{Introduction}
Understanding the behavior of gapless fermionic liquids in the vicinity of a quantum phase transition is at the heart of strongly correlated electron systems, dating back to the celebrated Hertz-Millis-Moriya framework~\cite{Hertz1976,Millis1993,Moriya1985}. In particular, the question of Fermi-liquid instabilities at a magnetic (quantum) phase transition~\cite{Stewart2001,Chubukov2004,Loehneysen2007,Chubukov2009,LiYi2014,ShenglongXu2015} and its applications to heavy-fermion materials~\cite{Custers2003,Steppke2013} and transition-metal alloys (such as cuprates and pnictides~\cite{LiuZhaoYu2016,ZhangWenLiang2016}), is of vital importance and broad interest to the %EBentire
condensed matter and high energy physics communities.

On the other hand, to be able to obtain the understanding of quantum critical phenomena in itinerant electron systems is extremely challenging.
Recently, extensive research efforts have been devoted to this question utilizing advanced renormalization group analysis, including the 2d Fermi surface coupled to a $U(1)$ gauge field, to Ising-nematic or to spin-density-wave bosonic fluctuations~\cite{Oganesyan2001,Metzner2003,Lee2009,Metlitski2010a,Metlitski2010b,Mross2010, Metlitski2015,Holder2015,Schlief2016}. Other approaches include dimensional regularization~\cite{Dalidovich2013,Lee2017} and working in the limit of a large number of boson flavors~\cite{Fitzpatrick2014}. Although important insights
have been revealed from these studies, many fundamental questions still remain open. For example, will anomalous dimensions arise at such an itinerant quantum critical point? The Hertz-Millis-Moriya theory predicts mean-field scalings. However, utilizing the effective field theory derived in Ref.~\cite{Metlitski2010a}, it was shown that divergence at four-loop order can lead to anomalous dimensions deviating from mean-field exponents~\cite{Holder2015}. In addition, the stability of these quantum critical points is also an open questions, e.g., will non-analyticities in the momentum and frequency dependence of the theory leads to an instability towards a fluctuation-induced first-order transition or a fluctuation-induced second-order transition~\cite{Kirkpatrick2003,Belitz2005}?
%is the critical point %EBis by itself
%stable, or does the transition always become first-order due to non-analyticities in the momentum and frequency dependence of the theory~\cite{Belitz2005}? 
Moreover, various studies have suggested that near such a quantum critical point, critical fluctuations may trigger some other instability, resulting in a new %EBquantum
phase that covers the QCP and masks %EBprevents us from accessing
the quantum critical region~\cite{Lederer2015}. Experimentally, a ferromagnetic QCP was reported in a heavy-fermion metal~\cite{Steppke2013}.
%\textcolor{red}{--Kai Can we move these two reference in this paragraph? Chubukov2004,Chubukov2009. They appeared very late in the manuscript, but its better to put them somewhere here.}

Since analytical approaches are facing difficulties to study itinerant QCPs in a controlled manner, here, we study such a QCP utilizing unbiased numerical calculations, i.e., the determinant quantum Monte Carlo (QMC) technique, which has been demonstrated as a very effective tool for such
problems~\cite{Berg12,Schattner2015a,Schattner2015b,ZXLi2015,Li2016925,Xu2016a,Lederer2016,Assaad2016,Gazit2016}.
For the numerical studies on QCPs in itinerant fermionic systems, one major challenge lies in the fact that critical quantum fluctuations often trigger strong effective attractions between fermions, resulting in instabilities in the particle-particle channel. For example, in the recent studies on Ising nematic and charge-density-wave (CDW) QCPs, superconducting domes are observed covering the QCPs~\cite{Berg12,Schattner2015a,Schattner2015b,ZXLi2015,Li2016925,Lederer2016}.
Although it is a very intriguing phenomenon that a QCP can induce unconventional superconductivity, for the study of quantum criticality itself, the induced superconducting dome makes it difficult to obtain direct information about the critical point for the following two reasons. (1) To measure accurately the critical exponents, it is important to examine the close vicinity of the QCP, which becomes challenging if the QCP is buried inside a new quantum phase. (2) It is known that quantum criticality is one source for non-Fermi liquid behaviors. In order to understand and characterize such a non-Fermi liquid induced by a QCP, it is important to suppress other ordering in the quantum critical region.
As a result, obtaining a pristine QCP without other induced instabilities becomes crucial for studying these QCPs, which is one main objective of this paper.

In this paper, we construct a model of 2d fermions interacting with gapless ferromagnetic (Ising) fluctuations, and %EBenineer a
use the determinant QMC technique to solve this (2+1)d problem exactly~\cite{Berg12,Schattner2015a,Schattner2015b,ZXLi2015,Li2016925,Xu2016a,Lederer2016,Assaad2016,Gazit2016}.
Our QMC results are consistent with such a pristine %EBconfirm a stable
continuous quantum phase transition in itinerant ferromagnet in (2+1)d, with
%In direct contrast with other similar QCPs, e.g. the Ising nematic or CDW, where the QCP is covered by a superconducting phase,
%our model shows
no superconducting ordering at any coupling strengths and down to the lowest temperature that we can access. The absence of superconductivity allows us to study the close vicinity of the QCP, where we found clear signatures of non-Fermi-liquid behavior
in the fermion propagators, induced by critical fluctuations. Furthermore, we find that due to the coupling between fermions and bosonic modes, the ferromagnetic QCP is %EBfound neither the naive bosonic one (i.e.,
different from an ordinary (2+1)d Ising transition~\cite{Pfeuty1971}, but also deviates from the predictions of the Hertz-Millis-Moriya theory~\cite{Loehneysen2007}.
Hence, our results support the existence of a ferromagnetic QCP with markedly non-Fermi liquid behavior. These results broaden the theoretical understanding of itinerant quantum criticality and make connections to the existing experiment phenomena.

\begin{figure*}
\includegraphics[width=\textwidth]{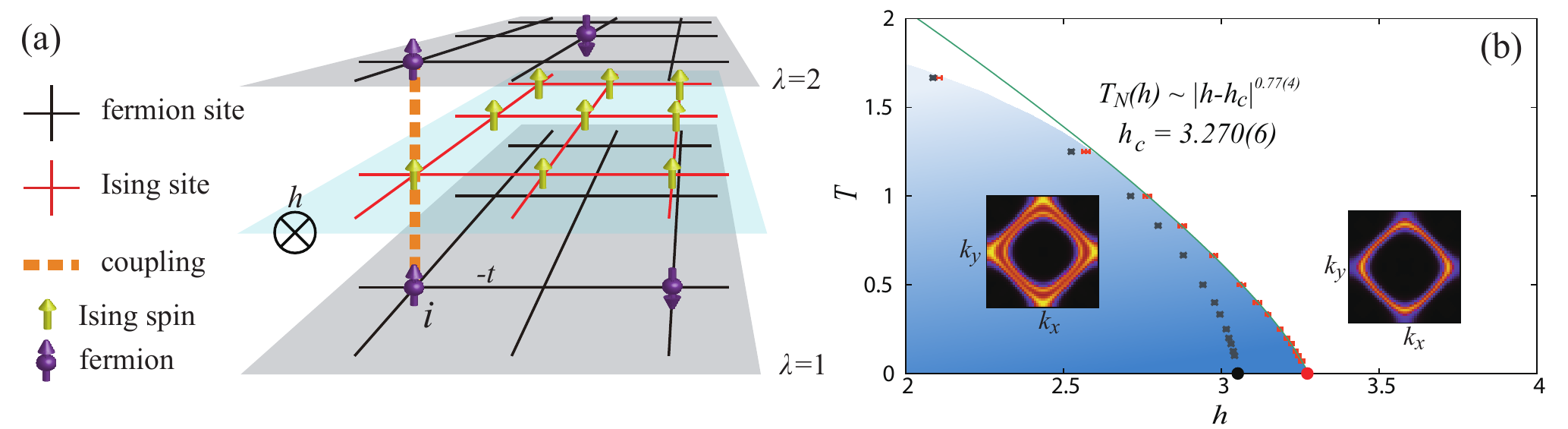}
\caption{(a) Sketches of our model shown in Eq.~\ref{eq:hamiltonian}. Two bands/orbits ($\lambda=1,2$) of fermions moving in a square lattice, at each site, the fermion spin is coupled to an Ising-gauge variable on the same lattice, the Ising spins are subject to a ferromagnetic interaction $J$ among themselves and a transverse magnetic field $h$. The quantum fluctuations of the transverse field Ising model furthermore introduce effective interactions among the fermions.
(b) $h-T$ phase diagram of our model in Eq.~\ref{eq:hamiltonian}. The phase boundary (orange data points) for coupling $\xi=1$ and $\mu=-0.5$ (fermion density $\langle n_{i\lambda}\rangle\approx0.8$) is the thermal transition points $T_{N}(h)$ from ferromagnetic phase (FM) to paramagnetic phase (PM). The FM phase is further highlighted by the shaded blue area. For comparison, we also plot the phase boundary without coupling (grey data points), i.e, that of the (2+1)d transverse field Ising model. The coupling changes the position of the QCP (black dot to red dot along the $h$-axis) as well as the power law of phase boundary, which is described by critical exponents $\nu$ and $z$ ($T_{N}(h)\sim |h-h_c|^{z \nu}$) with $\nu z=0.63$ for (2+1)d Ising at $\xi=0$ and $T_{N}(h) \sim |h-h_c|^{c}$ with $c=0.77(4)$ for FM-QCP at $\xi=1$.
The insets are plots of the low energy spectral weight, $G(\mathbf{k},\tau=\frac{\beta}{2})$, in the FM phase (left) and the PM phase (right), shown for L=24. Here, results from several different sets of twisted boundary conditions are superimposed, see Appendix \ref{app:flux}.}
\label{fig:model_and_phase_diagram}
\end{figure*}

\section{Model and Method}
%EBWe introduce a ferromagnetic quantum critical point to an itinerant system by coupling a Fermi liquid with
%a ferromagnetic transverse-field Ising model
We consider a two-dimensional lattice model of itinerant fermions coupled to an Ising ferromagnet with a transverse field [Fig.~\ref{fig:model_and_phase_diagram}(a)].
The Hamiltonian is comprised of three parts,
\begin{equation}
\hat{H}=\hat{H}_{f}+\hat{H}_{s}+\hat{H}_{sf}.
\label{eq:hamiltonian}
\end{equation}
\begin{eqnarray}
\hat{H}_{f}=-t\sum_{\langle ij\rangle\lambda\sigma}\hat{c}_{i\lambda\sigma}^{\dagger}\hat{c}_{j\lambda\sigma}+h.c.
%-\mu\sum_{i\lambda\sigma}\hat{n}_{i\lambda\sigma}
\\
\hat{H}_{s}=-J\sum_{\langle ij\rangle}\hat{s}_{i}^{z}\hat{s}_{j}^{z}-h\sum_{i}\hat{s}_{i}^{x}  \\
\hat{H}_{sf}=-\xi\sum_{i}s_{i}^{z}(\hat{\sigma}_{i1}^{z}+\hat{\sigma}_{i2}^{z})
\end{eqnarray}
The fermionic part, $\hat{H}_{f}$, describes spin-$1/2$ fermions on a square lattice with two
independent orbitals per site ($\lambda=1,2$). It includes a nearest-neighbor hopping term that preserves the spin and orbital sub-indices, where $i$ and $j$ labels the sites, while $\sigma$ and $\lambda$ are the spin and orbital indices, respectively.
We work in the grand canonical ensemble, where the fermion density %EBand the size of the Fermi sea
is set by
the chemical potential $\mu$. All energy scales are measured in units of $t$.
%where $\mu$ is the chemical potential.
In addition, each site of the square lattice has an Ising spin $\hat{s}_{i}^{z}=\pm 1$, %\textcolor{red}{(--Kai: Please check whether it is $\pm 1$ or $\pm 1/2$)}
whose quantum dynamics are governed by
a ferromagnetic transverse field Ising model $\hat{H}_{s}$.
The Ising spins and the fermions are coupled via an on-site Ising term
$\hat{H}_{sf}$,
where $\hat{\sigma}_{i\lambda}^{z}=(\hat{n}_{i\lambda\uparrow}-\hat{n}_{i\lambda\downarrow})/2$
is the $z$ component of the fermion spin at orbit $\lambda$ on site $i$.

While generically, models describing ferromagnetic transitions in fermionic systems suffer from the sign-problem,
here, the absence of the sign-problem is guaranteed by the introduction of the two orbitals $\lambda=1$ and $2$.
The two-orbital model is invariant under the anti-unitary symmetry $i\tau_y K$,
where $\tau_y$ is a Pauli matrix in the orbital basis and $K$ is the complex conjugation operator, and is thus sign-problem-free ~\cite{Wu2005}.
The details of the DQMC implementation are presented in Appendix~\ref{app:dqmc}.

The Hamiltonian in Eq.\eqref{eq:hamiltonian} possesses a $SU(2)\times SU(2)\times U(1)\times U(1)\times Z_2$ symmetry,
where the two $SU(2)$ symmetries are independent rotations in the orbital basis for spin up and spin down,
the two $U(1)$ symmetries correspond to conservation of particle number with spin up and spin down,
and the $Z_2$ symmetry interchanges spin up and spin down while flipping the Ising spins $s_z \rightarrow -s_z$.

As $h$ and $T$ are reduced, the system undergoes a paramagnetic-ferromagnetic (PM-FM) phase transition, spontaneously breaking the $Z_2$ symmetry.
In the absence of coupling between the Ising spins and the fermions, the transition belongs to the Ising universality class~\citep{BatrouniScalettar2011}.
However, in the presence of the coupling $\xi$ between the Ising spins and the fermions, the system becomes strongly correlated near the ferromagnetic QCP.
Here, we focus on the properties of this exotic itinerant paramagnetic-ferromagnetic transition.
%EBthe fermion spins undergo a PM-FM phase transition and the fluctuations of the Ising spins introduce effective fermion interactions, renders the system strongly correlated. For itinerant fermions,
%This allows us to explore the more exotic itinerant paramagnetic-ferromagnetic transition.

% We emphasize that this model differs from related models describing other classes of metallic QCPs studied recently via sign-problem-free QMC
% simulations, such as the CDW QCP, and in particular the Ising-nematic QCP~\cite{Berg12,Schattner2015a,Schattner2015b,ZXLi2015,Li2016925,Lederer2016}.
% In addition, as shown in Appendix~\ref{app:sc_theory}, our model also contains a different Ising-fermion coupling.
% For quantum phase transitions in fermionic systems, different numbers of fermion specifies and/or
% different fermion-boson couplings can change the universality of a quantum critical point, even if the symmetry breaking pattern remains the same.
% In our model, as will be shown below, these differences change the fate of the quantum critical region. The superconducting ordering observed
% in the vicinity of Ising nematic and CDW QCPs is suppressed near our FM QCP.

\section{Phase diagram}
%\label{sec:phasediagram}
%{\it Phase diagram}\,---\,
For $\xi=0$, the Ising spins are decoupled from the fermions and the phase transition is governed by the transverse-field Ising model.
The phase diagram is shown in Fig.~\ref{fig:model_and_phase_diagram}(b), where the phase boundary is marked by grey data points.
Our numerical studies confirm that the phase boundary ends at a QCP at $T=0$ with (2+1)d Ising universality class~\cite{BatrouniScalettar2011}. Near the QCP, the transition temperature follows the scaling relation,
$T_{N}(h)\sim|h-h_c|^{\nu z}$ with $h_c=3.04(2)$ and $\nu z=0.63$, consistent with the literature~\cite{Pfeuty1971,Xu2016a}.

We now study the itinerant PM-FM transition by turning on the coupling between the fermions and the Ising spins.
We begin by setting the coupling strength $\xi=1$ and chemical potential $\mu=-0.5$, which gives rise to a fermion density
$\langle n_{i\lambda} \rangle \approx 0.8$.
As shown in Fig.~\ref{fig:model_and_phase_diagram}(b),
%EBthe phase diagram remains qualitatively similar %EBlargely remains
turning on the coupling shifts FM phase boundary (orange data  points)
to higher values of $T$ and $h$.
At this coupling strength, down to the lowest temperature that we have accessed, $\beta=100$ ($T=0.01$), we observe no signature of any additional ordered phases near the QCP.
%EBcompeting orderings
%or intermediate phases.
We identify the finite-temperature ferromagnetic transition by a finite-size scaling analysis of spin susceptibilities, as explained in Appendix~\ref{app:thermalphasetransition}.
Extrapolation towards zero temperature indicates that the itinerant PM-FM quantum phase transition occurs at $h_c=3.270(6)$, and is of second order,
but the scaling behavior near the QCP
%near this quantum critical point
deviates strongly from  that of the (2+1)d Ising universality class. For example, as shown in Fig.~\ref{fig:model_and_phase_diagram} (b),
the transition temperature  $T_{N}$ scales as $T_{N}(h)\sim|h-h_c|^{c}$ with $h_c=3.270(6)$ and $c=0.77(4)$. Note that due to the itinerant nature of the QCP, the exponent $c$ is no longer expected to obey the relation
%EBfollow the Ising form, i.e.,
$c = z\nu$~\cite{Loehneysen2007}.

Due to the non-zero coupling $\xi$, the Fermi surface structure changes across the FM transition.
%EBthe fermion spins also undergoes PM-FM phase transition across the phase boundary, which is directly reflected in Fermi-surface configurations.
The fermionic low-energy spectral weight shown in the inset of Fig.~\ref{fig:model_and_phase_diagram}(b), extracted from the imaginary time Green's function $G(\tau=\beta/2)$~\cite{Trivedi1995, Schattner2015a}, reveals the location of Fermi surface.
%EBBy examining the fermionic Green's functions,
%we plotted the Fermi surfaces [insets of Fig.~\ref{fig:model_and_phase_diagram}(b)].
In the PM phase ($h>h_c$), fermions with up and down spins share the same Fermi surface, due to the spin degeneracy (Ising symmetry) of the Hamiltonian [right inset of Fig.~\ref{fig:model_and_phase_diagram}(b)]. This degeneracy is lifted in the FM phase ($h<h_c$) due to spontaneous symmetry breaking, and thus the Fermi surface splits [left inset of Fig.~\ref{fig:model_and_phase_diagram}(b)].

At the QCP, low-energy fermionic excitations near the Fermi surface become strongly coupled with the critical bosonic spin fluctuations,
offering an ideal platform for investigating the itinerant FM-QCP.
%Intensive efforts have been paid to this
%novel quantum critical system~\cite{Chubukov2004,Chubukov2009} but many open questions still remain.
%Even the stability of such a QCP is still under debate.
Below, we demonstrate that this QCP is stable down to low energy scales, and more interestingly, that it is characterized by a dramatic breakdown of Fermi liquid behavior.
%EBBelow, we show that with the help of the unbiased QMC simulation, we have not only established the stability of the critical point, but more interestingly, found that non-Fermi liquid behavior emerges at its vicinity.

\section{Non-Fermi liquid behavior}
\begin{figure}
\includegraphics[width=\columnwidth] {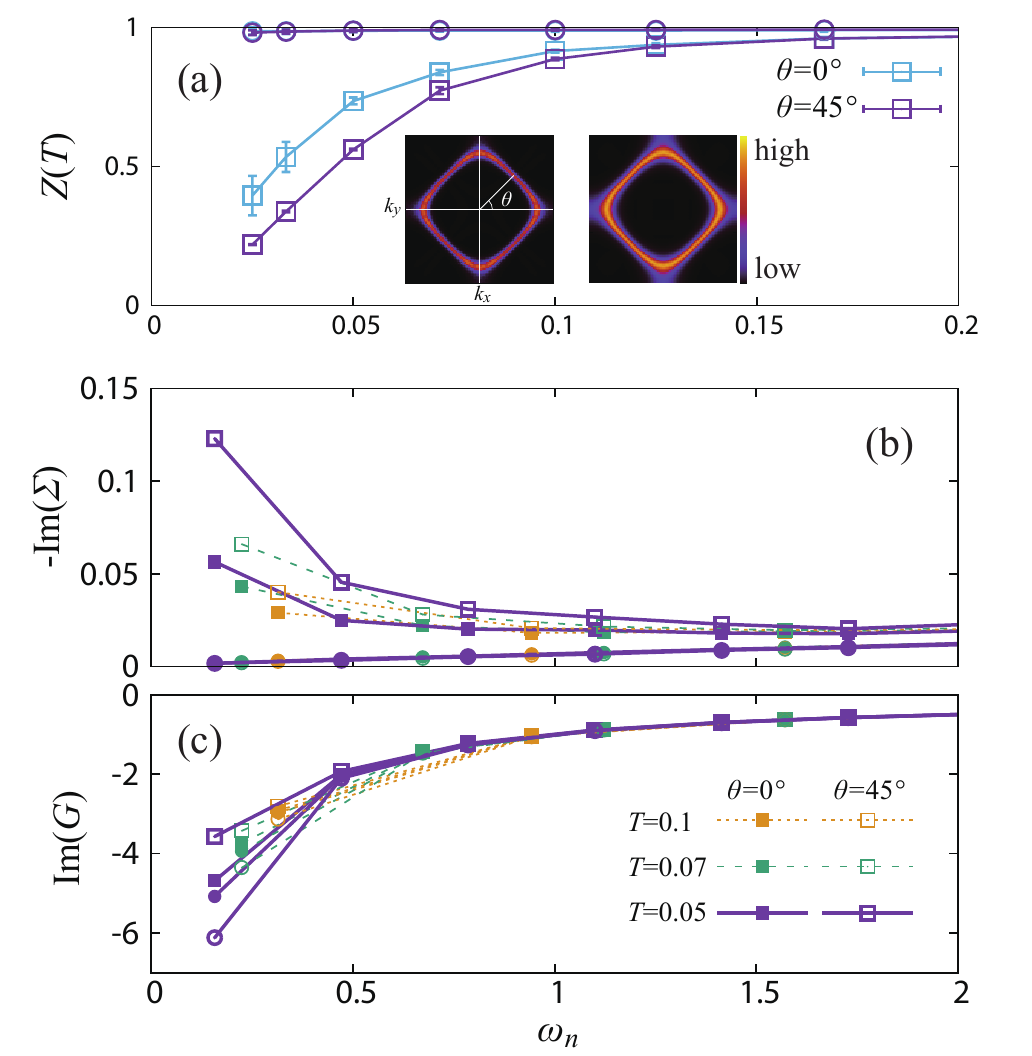}
\caption{(a) $Z_{\mathbf{k}_{F}}(T)$ at FM-QCP ($h_c=3.27$, squares) and in PM phase ($h=3.60$, circles). The left inset is the $G(\mathbf{k},\frac{\beta}{2})$ at FS for $T=0.05$ while the right inset is for $T=0.1$. Although there is anisotropy in $Z_{\mathbf{k}_{F}}$ at different parts of the FS, the quasi-particle weight in $k_x$ and $k_x=k_y$ directions both approach zero at FM-QCP, indicating a non-Fermi liquid behavior for the entire FS.  The data in the PM phase shows the quasi-particle weight approaching a constant (very close to 1), indicating the system is a Fermi liquid. (b) $-\text{Im}(\Sigma(\mathbf{k}_{F},\omega_n))$ at FM-QCP ($h=h_c$, square symbol), it increases as $\omega_n \to 0$ -- signifying the system at QCP loses their quasi-particle weight with a power law -- a non-fermi liquid behavior, while in PM phase ($h=3.60$, circle symbol) the imaginary part of self energy approaches zero linearly as $\omega_n \to 0$ -- a fermi liquid behavior.
(c) Imaginary part of the single-fermion Green's function at the FM-QCP  ($h=3.27$, square symbol) and in PM phase  ($h=3.60$, circle symbol). No signature of gap formation is observed.}
 %down to $T=0.025t$.}
\label{fig:zt_sig}
\end{figure}

%{\it Non-Fermi liquid behavior}\,---\,
%EBThe cleanness of the FM-QCP bestows us the possibility to directly probe the novel phenomena in the quantum critical region.
%The realization of an FM QCP opens the possibility to study its
One of the key theoretical prediction for an itinerant QCP is that the critical bonsonic fluctuations induce
strong damping of the fermions, resulting in a non-Fermi liquid where the low-temperature fermionic quasi-particle weight vanishes
at the Fermi surface~\cite{Oganesyan2001,Senthil2008,Lee2009,Metlitski2010a,Metlitski2010b,Dalidovich2013,Schlief2016,Lee2017}. Following~\cite{Chen2012}, we measure the quasi-particle weight from the Matsubara-frequency self-energy
%\begin{equation}
%Z_{\mathbf{k}_{F}}=\frac{1}{1-\left.\frac{\text{Im}\Sigma(\mathbf{k}_{F},i\omega_{n})}{\omega_{n}}\right|_{\omega_{n}\rightarrow0}}
%\label{eq:Z}
%\end{equation}
\begin{equation}
Z_{\mathbf{k}_{F}} \approx \frac{1}{1-\frac{\text{Im}\Sigma(\mathbf{k}_{F},i\omega_{0})}{\omega_{0}}}
\label{eq:Z}
\end{equation}
where $\Sigma$ is obtained from our QMC simulations, and $\mathbf{k}_{F}$ is the momentum at FS.  In our finite temperature simulations, the Matsubara frequencies take discrete values, $\omega_n = \pi (2n+1) T$. As an estimator for $Z_{\mathbf{k}_F}$, we use the first Matsubara frequency, $\omega_0 = \pi T$ in Eq.~(\ref{eq:Z}).

The results are shown in Fig.~\ref{fig:zt_sig}(a) for $h=h_c$ (squares) and $h>h_c$ (circles). Here, we plot the quasi-particle weight
at two different momentum points on the Fermi surface, i.e. $\vec{k}_F$ along the $k_x$ direction ($\theta=0$) and along the
$k_x=k_y$ direction ($\theta=\frac{\pi}{4}$). In the paramagnetic phase ($h>h_c$), $Z(T)$ remains close to unity at low temperature, indicating well-defined quasi-particles on the Fermi surface, as expected in a Fermi liquid. In contrast, at the QCP ($h=h_c$), $Z(T)$ is suppressed with decreasing temperature and extrapolates to
zero at $T\rightarrow 0$, which is the key signature of a non-Fermi liquid. This is one of the key findings of this study.

As shown in Fig.~\ref{fig:zt_sig}(a), at the QCP, as $T\rightarrow 0$, $Z(T)$ decreases faster for  $\vec{k}_F$ along the diagonal direction ($\theta=\frac{\pi}{4}$) in comparison with other parts of the FS (e.g. $\theta=0$). This anisotropy is due to the anisotropy of the Fermi surface. The fermion density in our simulation is around $0.8$, which is not far from perfect nesting (at half-filling). This near-nesting Fermi-surface  geometry results in soft fermion-bilinear modes around $\vec{Q}=(\pi,\pi)$, which we have directly observed by measuring the fermion spin susceptibility as shown in the Appendix~\ref{app:fermionsusceptibility}. These soft modes have a finite gap, and thus they are not the origin of the non-Fermi liquid behavior. However, scattering with these soft modes  presumably introduces additional dampings for the fermions. Such scattering processes are stronger
(weaker) for $\vec{k}_F$ near  the diagonal direction $\theta=\frac{\pi}{4}$ ($\theta=0$), where $2 \vec{k}_F$ is close to (far away from) $\vec{Q}$, and thus can lead to the observed anisotropy in $Z(T)$.

In Fig.~\ref{fig:zt_sig}~(b), we show the imaginary part of the self-energy, $\Sigma(\omega_n) = G_0^{-1} - G^{-1}$ (where $G(\omega_n)$ and $G_0(\omega_n)$ are the fermionic Green's function, and the Green's function of the non-interacting system, respectively) as a function of the Matsubara frequency $\omega_n$.
In the paramagnetic phase (circle symbols), $-\text{Im}(\Sigma)$ approaches zero linearly as $\omega_n \to 0$,
as expected for a Fermi liquid. Such a behavior is not seen at the QCP, however, where we observe an \emph{increase} of $-\text{Im}(\Sigma)$ upon decreasing $\omega_n$,
%EBas $\omega_n \to 0$, indiciating a power-law loss of quasi-particle weight upon lowering the temperature, which further confirms
%our observation of a non-fermi liquid
%~\cite{Senthil2008}.
indicating a strong damping of the fermions at low frequencies. One possible mechanism for such surprising frequency dependence is related to thermal fluctuations of the FM order parameter, as shown in appendix~\ref{app:self_energy}.

The fermion Green's functions, $\text{Im}G(\mathbf{k}_F,\omega_n)$, are presented in Fig.~\ref{fig:zt_sig}(c).
$-\text{Im}G(\mathbf{k}_F,\omega_n)$ is related to the single fermion spectral function $A(\mathbf{k}_F,\omega)$ through $-\mathrm{Im}G(\omega_n) = \int \frac{d\omega}{\pi} \frac{\omega_n}{\omega_n^2 + \omega^2}  A(\mathbf{k}_F, \omega)$.
%; in particular, when $\omega_n \to 0$, $\mathrm{Im}G(\omega_n) \rightarrow A(\omega=0)$.
The fact that $-\mathrm{Im}(G)$ increases with decreasing $\omega_n$ indicates that both for $h=h_c$ and $h>h_c$ and to the lowest temperature we have accessed, there is no visible suppression of the fermionic spectral weight at low frequency, \emph{i.e.} no sign of an opening of a gap in the fermionic spectrum (at least down to frequencies of the order of $\omega \sim \pi T$). A similar conclusion can be reached by noting that $-\mathrm{Im} \Sigma$ is never much larger than $\omega_0$, i.e the self-energy never dominates over the bare frequency dependence of the Green's function.
This observation  is consistent with the fact that there is no signature of a nearby superconducting phase for any value of $h$, as shown in Sec.~\ref{sec:superconductivity} and Appendix~\ref{app:superfluiddensity}. Similarly, there is no signature of any other competing phase that emerges close to the QCP (see Appendix \ref{app:fermionsusceptibility}).

%EBIn Fig.~\ref{fig:zt_sig} (c) and
%EBAppendix~\ref{app:superfluiddensity}, we show that although the $-\text{Im}(\Sigma)$ increases as $\omega_n \to 0$ at the FM-QCP, there is no sign of a gap opening in the fermionic spectrum %EBformation or spontaneous symmetry breaking
%in the vicinity of the QCP. In particular, the superfluid density measurement shows no signature of any superconductivity. Hence, we conclude that this non-fermi-liquid QCP remains robust even down to $T=0.025t$.

\section{Quantum critical scaling analysis}
%\label{sec:quantumcriticalanalysis}
%{\it Quantum critical scaling analysis}\,---\,%The non-Fermi liquid behavior at the FM-QCP also feed back to
Near the quantum critical point, the bosonic critical modes become strongly renormalized by the coupling to the gapless fermionic degrees of freedom.
As a result, the universality class of the quantum critical point is different from that of an ordinary Ising transition in $(2+1)$ dimensions.
Our QMC results indicate that the behavior at the QCP resembles the behavior predicted by Hertz-Mills theory, but also
deviates from it in a significant way. Below, we first summarize the Hertz-Millis predictions and then present a modified Hertz-Millis scaling formula, which fits our QMC data for the Ising spin susceptibility at all the momenta and frequencies simulated.

The Hertz-Millis-Moriya theory is based on quantum dynamics obtained from the random phase approximation
(RPA)~\cite{Hertz1976,Millis1993,Moriya1985}.
Within this approximation, the Ising spin susceptibility, $\chi(h,T,\vec{q},\omega_n) = \frac{1}{L^2}\int d\tau \sum_{ij} e^{i\omega_n\tau - i\mathbf{qr}_{ij}}\left \langle s_i^z(\tau) s_j^z(0)\right \rangle,$ takes the following form near the QCP,
\begin{align}
\chi(h,T,\vec{q},\omega_n)=\frac{1}{c_t T^2+c_h \left | h-h_{c} \right |+ c_q q^2+ c_\omega \omega^2+ \Delta(\vec{q},\omega_n)}.
\label{eq:HZ_chi}
\end{align}
where $c_t$, $c_h$, $c_q$, $c_\omega$ are constants. Here, the $c_q q^2+ c_\omega \omega^2$ comes from the bare action of the Ising
degrees of freedom, and the $\Delta(\vec{q},\omega_n)$ term is the contribution of the fermionic fluctuations.
For an isotropic 2D Fermi fluid, and for $q$ and $\omega_n$ much smaller than the Fermi momentum and energy, respectively,
\begin{align}
\Delta(\vec{q},\omega_n)=c_{\mathrm{HM}} \frac{|\omega_n|}{\sqrt{\omega_n^2+(v_f q)^2}}
\label{eq:Delta_RPA}
\end{align}
where $c_{\mathrm{HM}}$ is a constant and $v_f$ is the Fermi velocity. A key property of $\Delta(\vec{q},\omega_n)$ is its singular behavior in the limit
${q}\rightarrow 0$, $\omega \rightarrow 0$.
Depending on whether one first takes the long-wavelength ($q\to 0$) limit or the low-energy ($\omega \to 0$) limit,
$\Delta(\vec{q},\omega_n)$ converges to different values,
$\lim_{q\to 0}\lim_{\omega_n\to 0} \Delta(\vec{q},\omega_n)\sim \omega_n/q \to 0$, %EB (i.e., Hertz-Millis),
while $\lim_{\omega_n\to 0} \lim_{q\to 0}\Delta(\vec{q},\omega_n)=c_{\mathrm{HM}}$. %EB is finite.
This singularity is of great importance for the quantum dynamics at the QCP. It leads to the following property:
%on scaling relations
\begin{align}
\chi(h=h_c,T=0,\vec{q}, \omega_n=0)^{-1}&=c_q q^2
\\
\chi(h=h_c,T=0,\vec{q}=0, \omega_n)^{-1}&=c_{\mathrm{HM}}+c_{\omega} \omega_n^2
\end{align}
Beyond the RPA level, the scaling relation above can be modified by higher order terms, and the scaling analysis
in Ref.~\onlinecite{Millis1993} suggests that the exponent for $T$ shifts from $2$ to $1$ up to logarithmic corrections.

Although these two relations differ by a constant ($c_{\mathrm{HM}}$), note that the $q$  and $\omega_n$ %EBare expected to
dependence has the same scaling exponent. %EB, which is a nontrivial prediction and is not expected for typical non-itinerant QCPs.
The reason for this behavior %EBwhy $q$ and $\omega_n$ shares the same exponent here is because
is that at $q=0$ and small $\omega_n$ (or at $\omega_n=0$ and small $q$),
$\Delta(\vec{q},\omega_n)$ becomes independent of $q$ or $\omega_n$, respectively. Thus
the $\omega_n$ or $q$ dependence in $\chi^{-1}$ is dominated by the bare action of the Ising degree of freedom.
The fact that the exponents characterizing the $\omega$ and $q$ dependence are the same reflects the emergent Lorentz symmetry of the bare Ising action.
%EBBecause the low-energy effective theory of the Ising spins has an emergent Lorentz symmetry, $q$ and $\omega$
%shares the same scaling dimensions.

\begin{figure}
\includegraphics[width=\columnwidth]{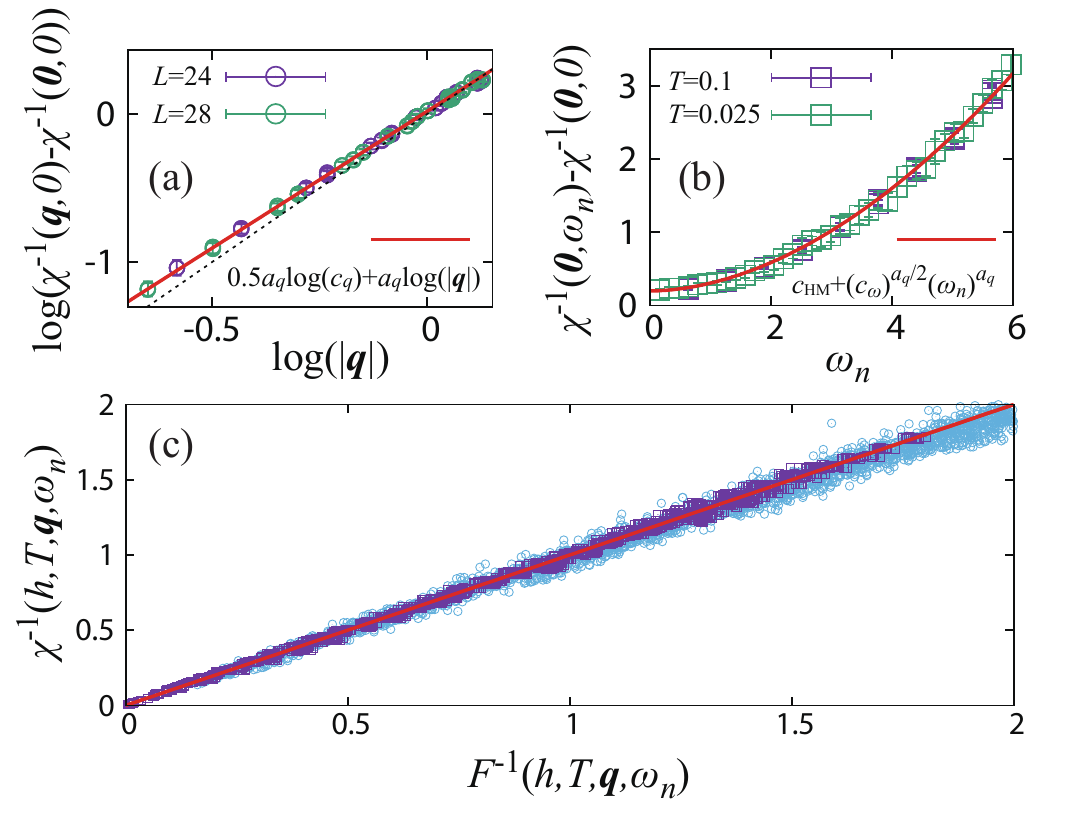}
\caption{
(a) Inverse spin susceptibility at $\omega_n=0$ as a function of $|\vec{q}|$ (data points with $L=24$, $28$ and $T=0.125$). The red line shows the fitting with $\chi^{-1}=c_q q^{a_q}$ and we get $a_q=1.85(3)$. The black dashed line shows the slope $a_q=2$.
(b) Inverse spin susceptibility at $q=0$ as a function of $\omega_n$ (data points with $L=20$, $24$ and $28$ for  $T=0.1$, with $L=20$ for $T=0.025$). The red curve shows the fitting with $\chi^{-1}=c_{\mathrm{HM}}+c_{\omega} \omega_n^{a_\omega}$.
(c) Data collapse for Ising susceptibility against the functional Eq.~\ref{eq:HZ_modified_chi}, where
$F^{-1}=c_t T^{a_t}+c_h \left | h-h_{c} \right |^{\gamma}+ \left(c_q q^2+ c_\omega \omega^2\right)^{a_q/2}
+ \Delta(\vec{q},\omega_n)$.
The dark violet square points (3946 in total, made up of $T=1.0,\  0.83,\  0.67,\  0.5,\  0.4,\  0.3,\  0.25,\  0.2,\  0.17,\  0.13,\  0.1$, $h=3.27,\  3.3,\  3.4,\  3.5,\  3.6,\  3.7,\  3.8,\  3.9,\  4.0$ and $L=24$) are for data with zero frequency while the light blue circle points (6096 in total, made up of $L=20$ with $T=0.1,\  0.05,\  0.033,\  0.025,\  0.014,\  0.01$, $L= 24,\ 28$ with $T=0.1,\  0.05$) are for data with frequency dependence. The red line with $y=x$ plays as the baseline for comparison MC data with functional approximation.}
\label{fig:wdependence}
\end{figure}

Our QMC results share some characteristics with this predicted form, but with anomalous scaling dimensions.
As shown in Figs.~\ref{fig:wdependence}(a) and (b), we indeed find that
$\lim_{q\to 0}\lim_{\omega_n\to 0} \chi^{-1}(\vec{q},\omega_n)$ differs from
$\lim_{\omega_n\to 0} \lim_{q\to 0}\chi^{-1}(\vec{q},\omega_n)$ by a constant $c_{\mathrm{HM}}=0.20(4)$
as predicted by the RPA. The ferromagnetic  susceptibility is found to be well-described by the following formula:

%EB$\vec{q}$ and $\omega$ do share the same scaling exponent, but its value is $1.85$.
%This anomalous dimension deviates from both the Ising exponent ($1.96$) observed at $\xi=0$ and the
%RPA  and Hertz-Millis prediction ($2$), which is a key result of our study.

%EBTo better incorporate quantum fluctuations and anomalous scaling dimensions at this QCP, we propose a
%modified formula
\begin{align}
&\chi(h,T,\vec{q},\omega_n) \nonumber
\\
&=\frac{1}{c_t T^{a_t}+c_h \left | h-h_{c} \right |^{\gamma}+ \left(c_q q^2+ c_\omega \omega^2\right)^{a_q/2}
+ \Delta(\vec{q},\omega_n)},
\label{eq:HZ_modified_chi}
\end{align}
with an anomalous exponent $a_q = 1.85(3)$ . This is different both from the exponent for an Ising transition in $(2+1)$ dimensions, $1.96$, and from the Hertz-Millis value of $2$. The presence of an anomalous exponent is another  key finding of this study.

The functional form in Eq.~(\ref{eq:HZ_modified_chi}) is analogous to the RPA prediction [Eq.~\eqref{eq:HZ_chi}], but allows for
non-mean-field exponents ($a_t$, $\gamma$ and $a_q$). %, whose values are determined by
%EBnumerical fitting.
This form is found to fit all the numerical data points
[Fig.~\ref{fig:wdependence}(c)].

Note that Eq.~\eqref{eq:HZ_modified_chi} contains $\Delta(\vec{q},\omega_n)$, which has the form of a free-fermion susceptibility. %EB, it will also receive modifications from quantum fluctuations, and the
%non-mean-field scaling dimensions shown above strongly suggest that the quantum dynamics shall deviates
%from Hertz-Millis scaling, which has $a_q=2$.
However, as will be shown below, within numerical resolution, as long as $\Delta$ captures
the singular behavior at $q\rightarrow 0$ and $\omega \rightarrow 0$, the quality of the fit of $\chi(h,T,\vec{q},\omega_n)$ is not sensitive to the detailed functional form of $\Delta(\vec{q},\omega_n)$.
More importantly, our key conclusions, e.g., the anomalous dimension $\eta=2-a_q=0.15$, are fully
independent of the particular choice of $\Delta(\vec{q},\omega_n)$.
Therefore, we use here the RPA form [Eq.~\eqref{eq:Delta_RPA}] where, for simplicity, we set $v_f$ to $2$.

By fitting with $\chi(h,T)$ at $q=0$ and $\omega_n=0$, we find that $c_t=0.13(1)$, $a_t=1.48(4)$, $c_h=0.7(1)$,
$\gamma=1.18(4)$ , details can be found in Appendix~\ref{app:isingsusceptibility}. As shown above, by fitting with $\chi$ at $q=0$ or $\omega=0$ [Figs.~\ref{fig:wdependence}(a) and (b)],
we find $a_q=1.85(3)$, $c_q=1.00(2)$, $c_{\omega}=0.10(2)$ and $c_{\mathrm{HM}}=0.20(4)$ .
With all the fitting parameters fixed, we can use Eq.~\ref{eq:HZ_modified_chi} to collapse all $\chi$ data
for all $q$, $\omega_n$, $h$ and $T$. The results are present in Fig~\ref{fig:wdependence}(c).
Clearly, all the data collapse onto the same curve, especially at small $q$, $\omega_n$, low temperature $T$ and $h\sim h_c$.

For $\Delta(\vec{q},\omega_n)$, we find that as long as $\lim_{\omega_n\to 0} \Delta(\vec{q},\omega_n)=0$
and $\lim_{q\to 0}\Delta(\vec{q},\omega_n)=c_{\mathrm{HM}}=0.20(4)$, modifying the functional form of
$\Delta(\vec{q},\omega_n)$ has little impact on $\chi$ within numerical error bars of the QMC data.
This uncertainty in $\Delta(\vec{q},\omega_n)$ prevents us from obtaining detailed information about the dynamics at the QCP (e.g., the precise value of the dynamical critical exponent $z$).
However, since by definition $\Delta(\vec{q},\omega_n)$ vanishes at $\omega_n=0$,
all the conclusions regarding the static limit ($\omega_n=0$), including the anomalous dimensions $\eta=2-a_q=0.15$
obtained from the $q$-dependence in $\chi(\vec{q},\omega_n=0)$
%EB, are robust and unbiased, fully independent of
are independent of the
details of $\Delta(\vec{q},\omega_n)$.

\section{Superconductivity}
\label{sec:superconductivity}
We now turn to discuss the superconducting properties close to the FM-QCP.
Unlike previous QMC studies of quantum criticality in metals, such as the Ising-nematic, SDW and other QCPs~\cite{Berg12,Schattner2015a,Schattner2015b,ZXLi2015,Li2016925,Lederer2016},
the FM-QCP shows a substantial separation of scales between the onset of superconducting correlations and the phenomena described thus far, i.e non-Fermi liquid behavior and quantum-critical scaling, thus making this system an ideal platform for the investigation of quantum fluctuations close to criticality.

As described in Appendix~\ref{app:sc_theory}, the spin fluctuations induce an attractive interaction in the spin-triplet channel.
The two-band structure of the model allows for a number of distinct superconducting order parameters, of which we find that the strongest pairing tendencies occur in the orbital-singlet, spin-triplet channel, with the order parameter
\begin{equation}
    \label{eq:SC_OP}
    \Delta_{i,\sigma} = c_{i1\sigma} c_{i2\sigma},
\end{equation}
where $\sigma=\uparrow,\downarrow$ is the spin index. Indeed, this channel is found to be the leading instability in a weak-coupling, mean-field analysis, see Appendix~\ref{app:sc_theory}.
The two components $\Delta_\uparrow$ and $\Delta_\downarrow$ are related by the $Z_2$ (spin-flip) symmetry of the model, and are therefore of equal magnitude in the magnetically disordered phase. The order parameter transforms as a scalar under lattice rotations and reflections, and hence we expect single-fermion excitations to be fully gapped in the superconducting state.

Because fermions in our model only preserve a $U(1)$ spin rotational symmetry,
%YS:although spin fluctuations in this triplet channel can reduce the superconducting transition temperature,
a finite-temperature triplet ordering is in principle allowed, in contrast to 2D systems with $SU(2)$ spin rotational symmetry where triplet pairing may only occur at zero temperature.  As shown in Appendix~\ref{app:exotic_sc}, the two-component nature of the order parameter allows, in principle, for more exotic phases such as a charge $4e$ superconductor and a spin-nematic phase~\cite{Kruger2002,Podolsky2009,Berg2009,Jiang2016}.
We have found no numerical evidence for the existence of these phases, and therefore focus on the triplet, charge $2e$ superconductor described above.

\begin{figure}
\includegraphics[width=\columnwidth]{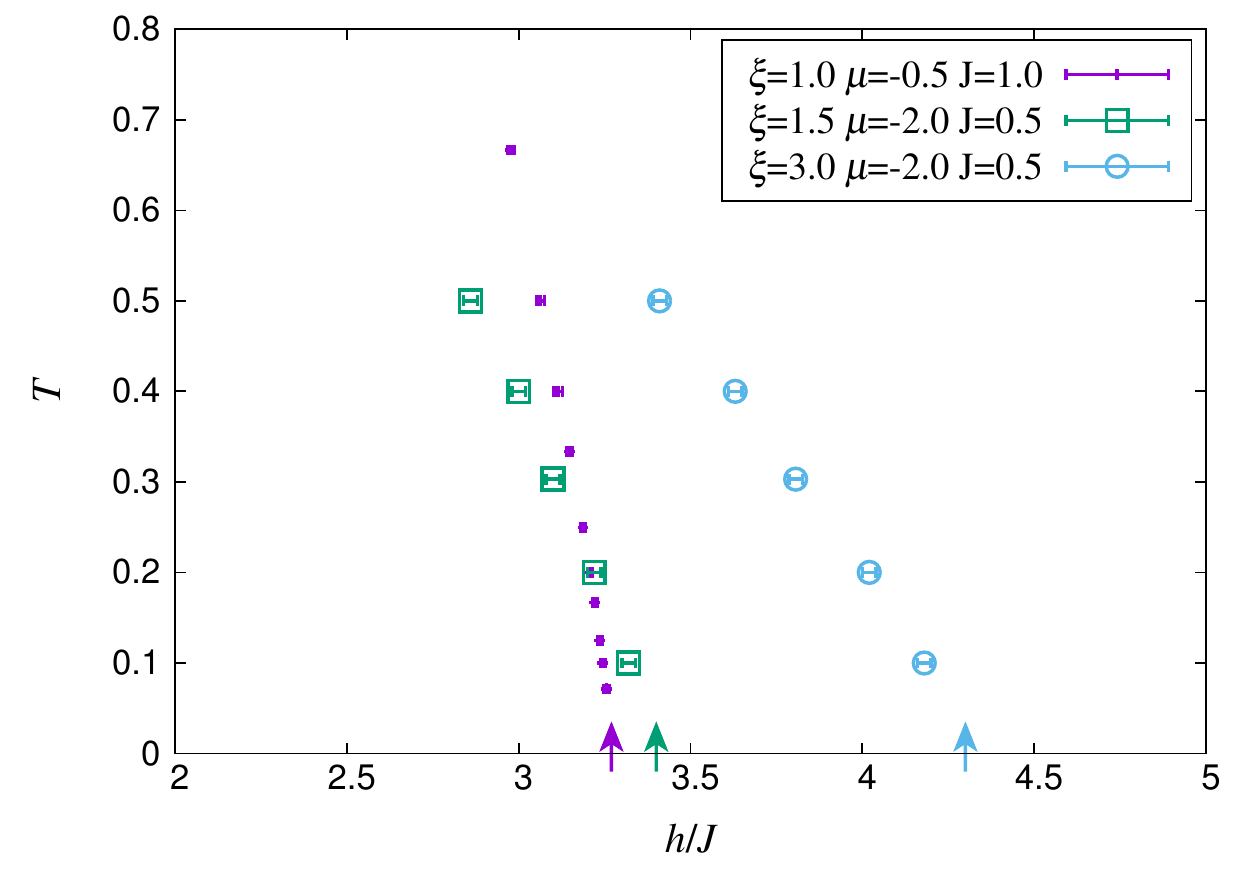}
\caption{Estimation of the FM-QCP for larger coupling. There are three parameter sets in total:  \{$\xi=1.0,J=1.0,\mu=-0.5$\}, \{$\xi=1.5,J=0.5,\mu=-2$\} and \{$\xi=3.0,J=0.5,\mu=-2$\}. The finite temperature phase boundaries are determined from the Ising spin susceptibilites, as discussed in Appendix \ref{app:isingsusceptibility}. The QCPs are denoted by the colored arrows. Close to each FM-QCP, we scan $h$ to measure various superconductivity instabilities. The results are shown in Fig.~\ref{fig:pair2all}.}
\label{fig:largexihN}
\end{figure}

To probe for superconducting tendencies near the QCP, we explore three parameter sets,
\{$\xi=1.0,J=1.0,\mu=-0.5$\}, \{$\xi=1.5,J=0.5,\mu=-2$\}, and \{$\xi=3.0,J=0.5,\mu=-2$\}.
First, the finite temperature FM to PM phase transitions are identified by finite-size scaling of the Ising spin susceptibilities,
as discussed in Appendix~\ref{app:isingsusceptibility}. Then, as depicted in Fig.~\ref{fig:largexihN}, the FM-QCPs are located by extrapolating the finite temperature
phase boundary towards $T=0$.
As expected, the larger the coupling $\xi$, the higher the critical field $h_c$ at the FM-QCP.

Next, we calculate the pairing correlations for all pairing channels, with on-site and nearest-neighbor form-factors (see %Tab.~\ref{tab:pairchanel} in
Appendix~\ref{app:superfluiddensity}).
Among all the available channels, only the order parameter defined in Eq.~\eqref{eq:SC_OP} has pairing correlations peaked at the QCP. We conclude that it is the only channel which is substantially enhanced by critical FM fluctuations.

\begin{figure}
\includegraphics[width=\columnwidth]{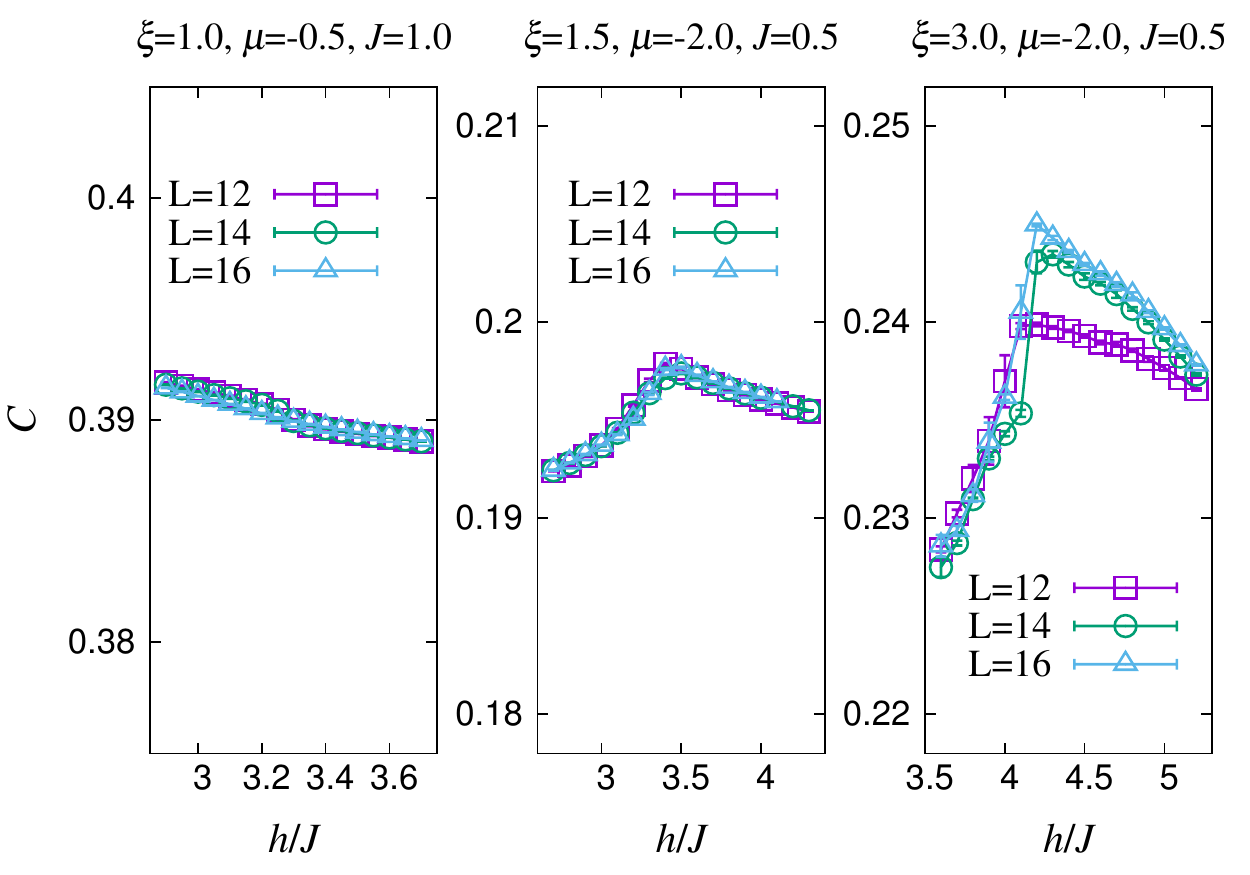}
\caption{Static pairing-correlation function $C=\frac{1}{L^{2}}\langle\hat{\Delta}^{\dagger}\hat{\Delta}\rangle$
for order parameters defined in Eq.~\eqref{eq:SC_OP}.
For \{$\xi=1.0,J=1.0,\mu=-0.5$\}, no enhancement of pairing correlation functions is observed in any pairing channel down to $T=0.025$.
For \{$\xi=1.5,J=0.5,\mu=-2$\} and \{$\xi=3.0,J=0.5,\mu=-2$\}, the pairing order parameters $\Delta_\uparrow$ and $\Delta_\downarrow$ show enhanced
correlation near the QCP, in agreement with theoretical analysis. No enhancement is observed in other pairing channels.
}
\label{fig:pair2all}
\end{figure}

Fig.~\ref{fig:pair2all} shows the pairing structure factor $C=\frac{1}{2 L^{2}}\sum_{ij\sigma}\langle \Delta_{i\sigma}^{\dagger} \Delta_{j\sigma}\rangle$.
At all couplings we find a peak of the pairing correlations at the QCP. At $\xi=1$ and $\xi=1.5$, the pairing correlations do not grow with the system size,
indicating we are far from a superconducting transition. At the strongest coupling, $\xi=3$, $C$ grows, albeit very slowly, with $L$, indicating that the correlation length of the superconducting order parameter is moderately large.

Although pairing correlations in this channel are enhanced close to the QCP, long-range or quasi-long range superconducting order never
develops down to $T=0.025$ for any of the three parameter sets.
This conclusion in corroborated by an analysis of the superfluid density, shown in Appendix~\ref{app:superfluiddensity}.

\section{Discussion}
%{\it Discussion}\,---\,
%EBWith the fermion couple to ferromagnetic fluctuated Ising field, we
%generate a clean QCP in an itinerant Fermi system.
In this work, we have constructed a lattice model that realizes a clean ferromagnetic QCP in an itinerant Fermi system.
At the QCP, clear non-Fermi liquid behavior is observed.
The static ($\omega_n=0$) ferromagnetic susceptibility obeys scaling with an anomalous
dimension $\eta=2-a_q=0.15$, which %EBstrongly
deviates from both the (2+1)d-Ising value ($\eta=0.036$) and the mean-field value expected from Hertz-Millis theory ($\eta=0$).

The  $\omega_n$ dependence of the bosonic susceptibility %EB, quantum dynamics near the QCP
is highly nontrivial. %EB and complicated.
In particular, the long-wave length limit ($q\rightarrow 0$) and
the low-frequency limit ($\omega_n \rightarrow 0$) do not commute.
To fully characterize the quantum dynamics at the QCP would requires to obtain detailed information about $\Delta(\vec{q},\omega_n)$
in Eq.~\eqref{eq:HZ_modified_chi}, especially in the low-energy limit. The presence of an anomalous dimensions in $a_q$ and in transition temperature $T_N$
suggest that $\Delta$ probably also deviates from the Hertz-Millis form, which has a dynamic critical exponent $z=3$ and thus
predicts mean-field exponents. At the qualitative level, our conclusions and the observation of anomalous dimensions are in good agreement with the four-loop scaling analysis in Ref.~\cite{Holder2015}.
However, although our studies reveal direct information on the qualitative features
of $\Delta(\vec{q},\omega_n)$ (i.e. the singular behavior in the limit $q\rightarrow 0$, $\omega_n \rightarrow 0$), the detailed functional form is beyond the resolution set by finite-size and finite-temperature effects of our QMC study.
%EBwhich requires further investigations.

It is interesting to contrast the results of the present study with those obtained for a related (but different) problem of an itinerant Ising-nematic QCP~\cite{Schattner2015a,Lederer2016}. The two problems have essentially the same description within Hertz-Millis theory.
Similarly to our results for an Ising ferromagnetic QCP, the Ising nematic QCP displays strong deviations from Fermi liquid behavior. A key difference, however, is that in the FM-QCP, Fermi-liquid behavior is lost everywhere along the Fermi surface, i.e there are no `cold-spots' with long-lived quasi-particles. Furthermore, near the FM-QCP, strong deviations from Fermi-liquid behavior were found at temperatures much larger than any scale associated with superconductivity.
In the Ising-nematic problem, the Ising order parameter correlations also show a singular behavior in the  $(q\rightarrow0, \omega_n\rightarrow 0)$ limit. However, in the FM case presented here, the fermionic magnetic susceptibility is exactly conserved, unlike the Ising-nematic case where an approximate conservation was found. An additional key difference is the anomalous exponents detected in the present work.

Perhaps the most striking difference between the two models lies in their superconducting properties. While the Ising-nematic QCP was found to be strongly unstable towards s-wave superconductivity~\cite{Lederer2016}, the leading superconducting instability in the FM case is described by a two-component, spin triplet order parameter, with the possibility of exhibiting interesting, unconventional phases such as a charge 4e superconductor.

On the experimental side, our results may shed some new light on the study about 2d or quasi-2d metallic materials with Ising magnetic order, like Fe$_{1/4}$TaS$_2$~\cite{Morosan2007} and CeCd$_3$As$_3$~\cite{Liu2016}.

%Since the two problems have essentially the same description within Hertz-Millis theory, the differences may provide guides for the

\section*{Acknowledgments}
%{\it Acknowledgments}\,---\,
The authors thank F. Assaad and S. Kivelson for helpful discussions. XYX and ZYM are supported by  the Ministry of Science and Technology of China under Grant No. 2016YFA0300502, the National Science Foundation of China under Grant Nos. 11421092 and 11574359, and the National Thousand-Young-Talents Program of China. We thank the following institutions for allocation of CPU time: the Center for Quantum Simulation Sciences in the Institute of Physics, Chinese Academy of Sciences; the Tianhe-1A platform at the National Supercomputer Center in Tianjin; and the Gauss Centre for Supercomputing e.V. (www.gauss-centre.eu) for providing access to the GCS Supercomputer SuperMUC at Leibniz Supercomputing Centre (LRZ, www.lrz.de). KS is supported by the National Science Foundation under grants PHY-1402971 at the University of Michigan and the Alfred P. Sloan Foundation. YS and
EB were supported by the Israel Science Foundation under
Grant No. 1291/12, by the US-Israel BSF under Grant No.
2014209, and by a Marie Curie reintegration grant. EB was
supported by an Alon fellowship. YS and EB thank S.~Lederer and S.~Kivelson for a collaboration on related topics.

\appendix

\section{Determinantal quantum Monte Carlo implementations}
\label{app:dqmc}
The determinantal quantum Monte Carlo (DQMC) formalism starts with the partition function of original Hamiltonian. To efficiently evaluate the trace in partition function, discretized imaginary time is used and $\beta=M\Delta \tau $ ($\Delta \tau=0.05$). As the studied model contains both fermion and Ising degrees of freedom, the trace will involve both a sum over Ising spin configurations and a determinant after tracing out the fermion degrees of freedom.
\begin{equation}
\begin{split}
Z & =  \text{Tr}\left[e^{-\beta\hat{H}}\right]\\
& =  \sum_{s^{z}_{1}\cdots s^{z}_{N}=\pm1}\text{Tr}_{F}\left\langle s^{z}_{1}\cdots s^{z}_{N}\left|\left(e^{-\Delta\tau\hat{H}}\right)^{M}\right|s^{z}_{1}\cdots s^{z}_{N}\right\rangle
\end{split}
\end{equation}
Let $\vec{S}=\left(s^{z}_{1}\cdots s^{z}_{N}\right)$ denoting the Ising spins, then
\begin{equation}
\begin{split}
Z & = \sum_{\vec{S}_{1}\cdots\vec{S}_{M}}\text{Tr}_{F}\langle\vec{S_{1}}|e^{-\Delta\tau\hat{H}}|\vec{S}_{M}\rangle\langle\vec{S_{M}}|e^{-\Delta\tau\hat{H}}|\vec{S}_{M-1}\rangle \\
&\qquad\qquad \cdots \langle\vec{S_{2}}|e^{-\Delta\tau\hat{H}}|\vec{S}_{1}\rangle,
\end{split}
\end{equation}
now we can trace out the fermion degrees of freedom, and obtain the configurational weight,
\begin{equation}
\omega_{\mathcal{C}}=\omega_{\mathcal{C}}^{TI}\omega_{\mathcal{C}}^{F}
\end{equation}
with the Ising part
\begin{equation}
\omega_{\mathcal{C}}^{TI} =  \left(\prod_{\tau}\prod_{\langle i,j\rangle}e^{\Delta\tau Js^{z}_{i,\tau}s^{z}_{j,\tau}}\right)\left(\prod_{i}\prod_{\langle\tau,\tau'\rangle}\Lambda e^{\gamma s^{z}_{\tau,i}s^{z}_{\tau',i}}\right)
\end{equation}
where $\Lambda^{2}=\sinh(\Delta\tau h)\cosh(\Delta\tau h)$, $\gamma=-\frac{1}{2}\ln\left(\tanh(\Delta\tau h)\right)$.
For the fermion part, we have
\begin{equation}
\omega_{\mathcal{C}}^{F} =  \det\left(\mathbf{1}+\mathbf{B}_{M}\cdots\mathbf{B}_{1}\right)
\end{equation}
As an anti-unitary symmetry $i\tau_yK$ (where $\tau_y$ is a Pauli matrix in the orbital basis and $K$ is the complex conjugation operator)
 make the Hamiltonian invariant, the
fermion part ratio can be further rewritten as
\begin{equation}
\omega_{\mathcal{C}}^{F}=\left|\prod_{\sigma}\det\left(\mathbf{1}+\mathbf{B}_{M}^{1\sigma}\cdots\mathbf{B}_{1}^{1\sigma}\right)\right|^{2}
\end{equation}
where
\begin{equation}
\mathbf{B}_{\tau}^{\lambda\sigma}=\exp\left(-\Delta\tau\mathbf{K}^{\lambda\sigma}+\Delta\tau\xi\text{Diag}(s_{1}^{z},\cdots,s_{N}^{z})\right)
\end{equation}
with $\mathbf{K}^{\lambda\sigma}$ the hopping matrix for orbital $\lambda$ and spin $\sigma$. It turns out both the fermion weight and the Ising weight are always positive, thus there is no sign problem. To systematically improve the simulation, especially close to (quantum) critical point, we have implemented both local update in DQMC and space-time global update~\cite{Xu2016a}. In the global update, we use Wolff algorithm~\cite{Wolff1989} to propose space-time clusters of the Ising spins and then calculate the fermion weight to respect the detail balance as the acceptance rate of the update. Further attempts, with the recently developed self-learning determinantal quantum Monte Carlo scheme~\cite{Liu2016self,Liu2016fermion,Xu2016self}, which can greatly reduce the autocorrelation at (quantum) critical point and speedup the simulation with $\mathcal{O}(N)$ fold, be able to access larger $L$ and lower $T$, are in progress.

\section{z-direction flux}
\label{app:flux}
To reduce spurious finite size effects, we have used the techniques introduced in Ref.~\cite{AssaadEvertz2008}. The basic idea is to introduce an effective $z$-direction flux by multiplying the hopping parameter by Peierls phase factors, i.e
\[
\label{eq:flux_hopping}
-t \hat{c}_{i\lambda\sigma}^{\dagger}\hat{c}_{j\lambda\sigma} \rightarrow -te^{i\phi_{ij}^{\lambda\sigma}} \hat{c}_{i\lambda\sigma}^{\dagger}\hat{c}_{j\lambda\sigma},\]
with
 $\phi_{ij}^{\lambda\sigma} = \frac{2\pi}{\Phi_0} \int_{\vec{r}_i}^{\vec{r}_j}\vec{A}^{\lambda\sigma}(\vec{r})\cdot d\vec{r}$.

To make sure the model remains free of the sign problem, we take
\begin{equation}
    \label{eq:flux_signs}
\phi_{ij}^{1\uparrow} = \phi_{ij}^{1\downarrow} = - \phi_{ij}^{2\uparrow} = -\phi_{ij}^{2\downarrow},
\end{equation}
therefore the applied flux is not a true magnetic field, as it couples differently to fermions of different flavors.

Translational invariance imposes restrictions on the magnitude of the effective magnetic field, namely,
$B^{\lambda\sigma}=n\frac{\Phi_0}{L^2}$ with $n$ an integer and $\Phi_0$ the flux quanta.
We use the Landau gauge $\vec{A}^{\lambda\sigma}(\vec{r}) = -B^{\lambda\sigma}y\hat{\vec{x}}$ in the bulk of the system,
whereas special care is taken at the 'edges'
\begin{equation}
\phi_{i(x,y=L);j(x,y=1)}^{\lambda\sigma} = \frac{2\pi} {\Phi_0} B^{\lambda\sigma} L x
\end{equation}
\begin{equation}
\phi_{i(x,y=1);j(x,y=L)}^{\lambda\sigma} = -\frac{2\pi} {\Phi_0} B^{\lambda\sigma} L x.
\end{equation}

While applying a flux in the $z$ direction dramatically improves the convergence to the thermodynamic limit,
it also breaks translation symmetry, making it impossible to extract the momentum dependence of fermionic correlations.
Whenever such information is needed (such as for the momentum resolved, singlet-particle Green's function $G_{\mathbf{k}}(\omega_n)$, we apply a flux in the $x$ or $y$ direction, which is equivalent to twisting the boundary conditions. Just as in the case of the $z$ directed flux, choosing \eqref{eq:flux_signs} ensures the absence of the sign-problem \cite{Gerlach2017}. We have used six kinds of twisted boundary conditions, $(0,0)$, $(\pi/2,0)$, $(\pi/2,\pi/2)$, $(\pi,0)$, $(\pi,\pi/2)$ and $(\pi,\pi)$, thus to get 16 times higher resolution of momentum.

\section{Thermal phase transition}
\label{app:thermalphasetransition}
The thermal phase boundary in Fig.~\ref{fig:model_and_phase_diagram} (b) of the main text is controlled by the 2d Ising critical exponents $\gamma=7/4$ and $\nu=1$,
implying that the zero frequency and zero momentum Ising spin susceptibility around the finite temperature critical field $h_N$ satisfies
\begin{equation}
\chi(h,T,0,0)=L^{\gamma / \nu } f( (h-h_N) L ^{1/\nu} ).
\label{eq:datacollapse}
\end{equation}
Here, the Ising spin susceptibility is defined as
\begin{equation}
\chi(h,T,\vec{q},i\omega_{n})=\frac{1}{L^{2}}\sum_{ij}\int_{0}^{\beta}d\tau e^{i\omega_{n}\tau-i\vec{q}\cdot\vec{r}_{ij}}\langle s^{z}_{i}(\tau)s^{z}_{j}(0)\rangle,
\end{equation}

\begin{figure}
\includegraphics[width=\columnwidth]{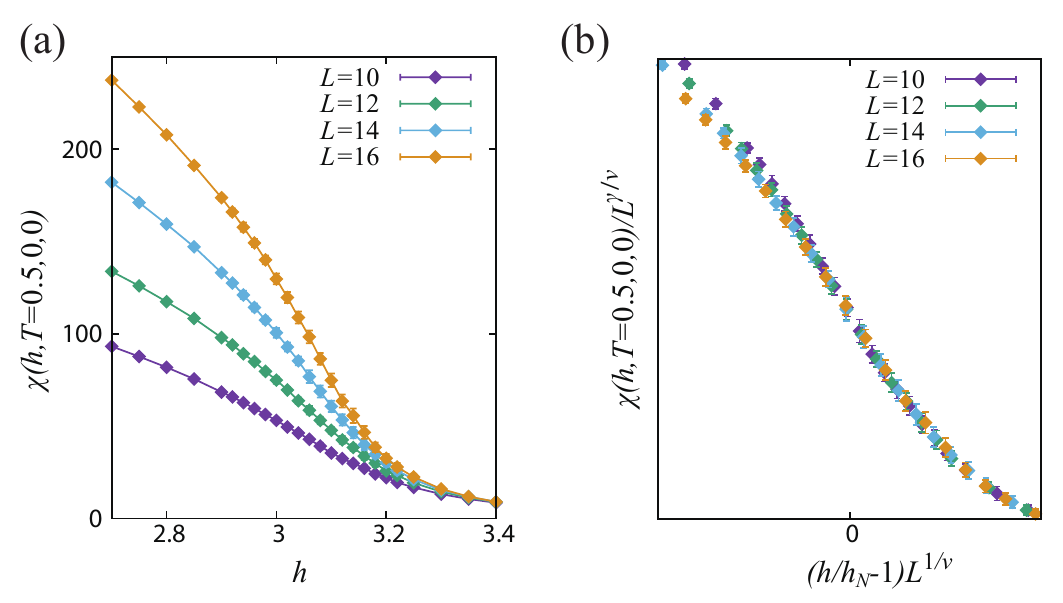}
\caption{Finite temperature FM to PM phase transition at $T=0.5$. (a) shows the Ising spin susceptibility at $\mathbf{q}=0$ and $\omega_n =0$, and (b) shows the data collapse according to Eq.~\ref{eq:datacollapse}. $h_{N}$ is a free fitting parameter and the best data collapse gives $h_N(T=0.5)\approx 3.06$. }
\label{fig:gethn_s}
\end{figure}

\begin{figure}
\includegraphics[width=\columnwidth]{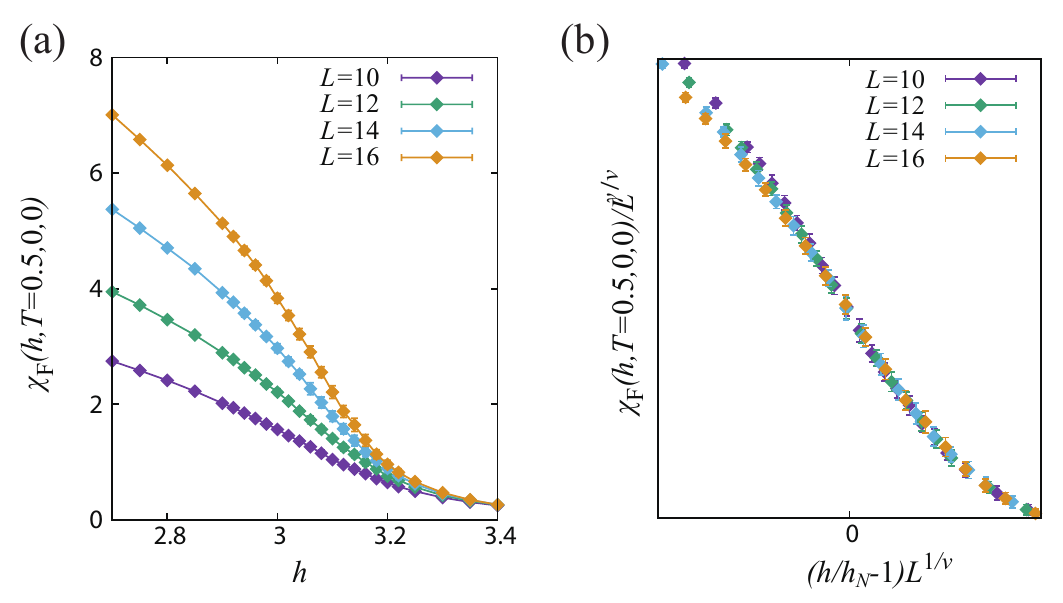}
\caption{Finite temperature FM to PM phase transition at $T=0.5$. (a) shows the fermion spin susceptibility at $\mathbf{q}=0$ and $\omega_n =0$, and (b) shows the data collapse according to Eq.~\ref{eq:datacollapse}. $h_{N}$ is a free fitting parameter and the best data collapse gives $h_N(T=0.5)\approx 3.06$. }
\label{fig:gethn_f}
\end{figure}

Fig.~\ref{fig:gethn_s} (a) illustrates the behavior of $\chi(h,T,0,0)$ at fixed temperature $T=0.5$ as a function of transverse field, Fig.~\ref{fig:gethn_s} (b) is the data collapse according to Eq.~\ref{eq:datacollapse}, from which we can obtain $h_N(T=0.5)\approx3.06$. Due to the coupling between the fermions and the Ising spins, the fermions go through the same finite temperature phase transition, as shown in Fig.~\ref{fig:gethn_f} (a). The fermionic spin susceptibility $\chi_\mathrm{F}(h,T,0,0)$, where
\begin{equation}
\label{eq:fermionspinsuscep}
     \chi_\mathrm{F}(h,T,\vec{q},\omega_n) = \frac{1}{L^2}\int d\tau \sum_{ij\lambda\lambda'} e^{i \omega_n\tau-i\mathbf{q}\mathbf{r}_{ij}}\left \langle \sigma_{i\lambda}^z(\tau) \sigma_{j\lambda'}^z(0)\right \rangle,
    \end{equation}
behaves much like $\chi(h,T,0,0)$, and a data collapse, shown in Fig.~\ref{fig:gethn_f} (b) gives rise to the same $h_N(T=0.5)\approx3.06$.

\section{Quantum critical scaling analysis of $\chi(h,T,\vec{0},0)$}
\label{app:isingsusceptibility}
In the main text, we have discussed the dynamic Ising spin susceptibility, $\chi(h,T,\vec{q},\omega_n)$, and performed the quantum critical scaling analysis. Here we reveal more details.

\begin{figure}
\includegraphics[width=\columnwidth]{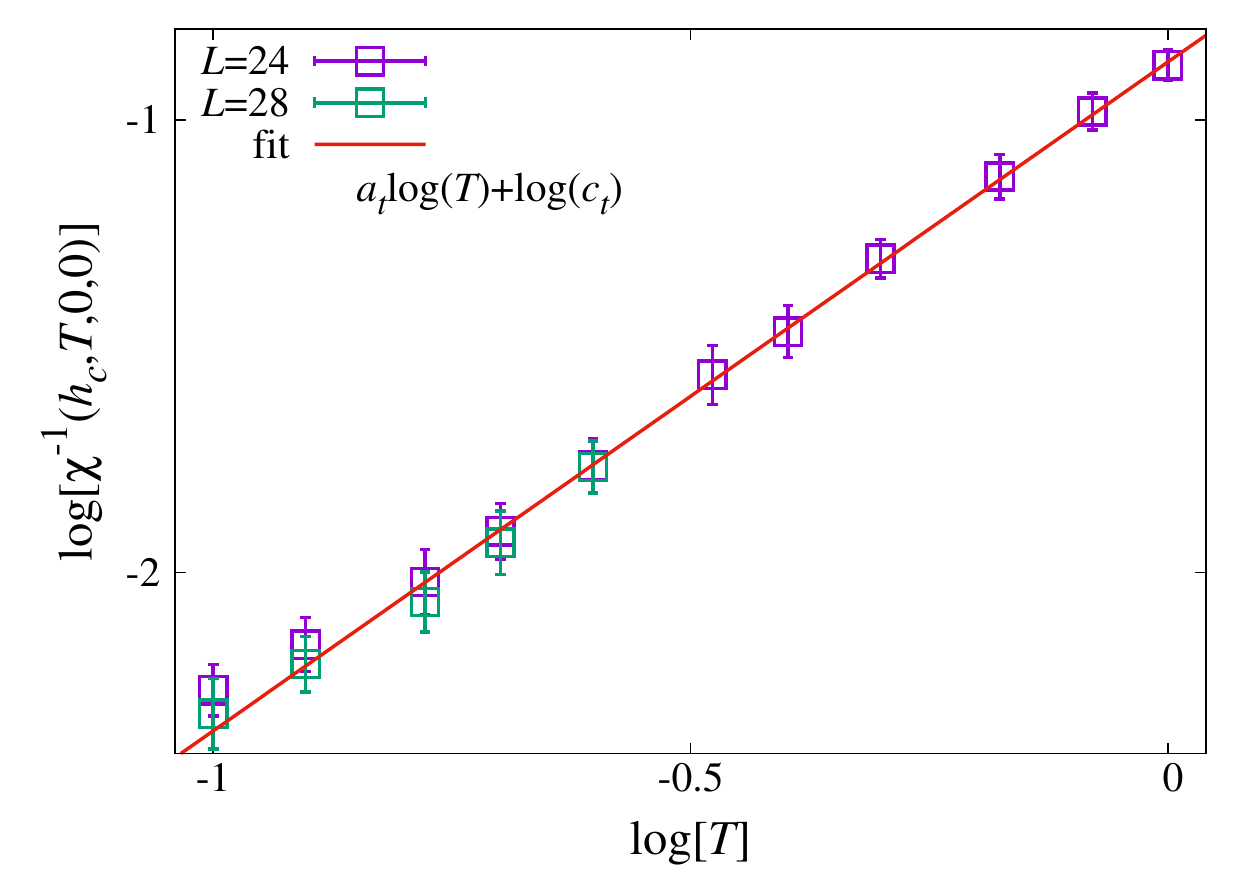}
\caption{Inverse Ising spin susceptibility at QCP ($\chi^{-1}(h_c,T,\mathbf{0},0)$) as a function of temperature $T$, the slope of the log-log plot reveals the power $a_t=1.48(4)$ and the intercept gives rise the prefactor $c_{t}=0.13(1)$.}
\label{fig:logchivslogT}
\end{figure}

\begin{figure}
\includegraphics[width=\columnwidth]{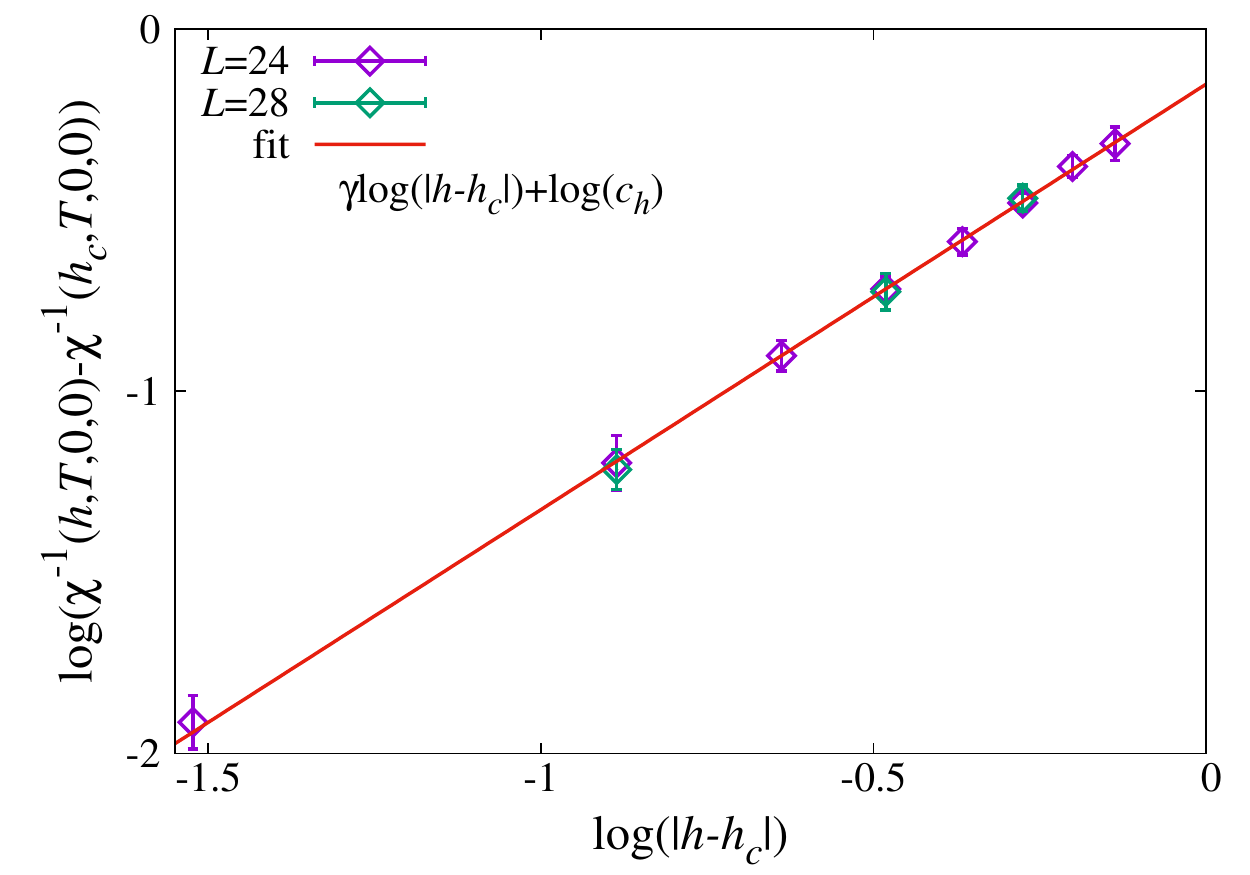}
\caption{Inverse Ising spin susceptibility ($\chi(h,T=0.1,\mathbf{0},0)$) away from QCP ($h>h_c$) as a function of $|h-h_c|$.  Here the temperature is $T=0.1$ and $\chi^{-1}(h_c,T=0.1,\mathbf{0},0)$ has been deducted in the plot. The slope of the log-log plot reveals the power $\gamma=1.18(4)$ and the intercept gives rise the prefactor $c_{h}=0.7(1)$.}
\label{fig:logchivslogh}
\end{figure}

According to Eq.~\ref{eq:HZ_modified_chi} in the main text, at $\vec{q}=\vec{0}$ and $\omega_n =0$, we have
\begin{equation}
\chi(h,T,\mathbf{0},0)=\frac{1}{c_t T^{a_{t}}+c_h |h-h_{c}|^{\gamma}}.
\end{equation}
We detect the power in $T$ by measuring $\chi(h=h_c,T,\vec{0},0)$ and Fig.\ref{fig:logchivslogT} shows the fitting of $\chi(h=h_c,T,\vec{0},0)^{-1}=c_t T^{a_t}$. The fit gives rise $c_{t}=0.13(1)$ and $a_{t}=1.48(4)$. We further detect the power in transverse field $h$ by fitting $\chi(h,T,\vec{0},0)^{-1} - \chi(h_c,T,\vec{0},0)^{-1}=c_h |h-h_c|^\gamma$, showed in Fig.~\ref{fig:logchivslogh}. The fit gives rise $c_h=0.7(1)$ and $\gamma=1.18(4)$.

\section{Effective attraction and BCS analysis}
\label{app:sc_theory}
In this section, we examine the effective attraction induced by the Ising-field fluctuations.
Away from the critical point, the Ising spins are gapped, and we can therefore integrate them out, obtaining an effective four-fermion interaction,
\begin{align}
S_{\textrm{int}}= -\frac{\xi^2}{2} \int d\tau d\tau' \sum_{i j\lambda \lambda'} \sigma^z_{i\lambda}(\tau)  \chi(h,T,\mathbf{r}_i-\mathbf{r}_j,\tau-\tau') \sigma^z_{j\lambda'}(\tau').
\end{align}
This interaction contains two types of terms: (a) attractions, e.g $n_{\lambda\uparrow}n_{\lambda'\uparrow}$, (b) repulsions, $n_{\lambda\uparrow}n_{\lambda'\downarrow}$.
Focusing on instabilities in the particle-particle channel, only attractive interactions will be considered. We are therefore restricted to pairing of fermions of equal spin.
Within the BCS approximation, to gain the most free energy, a nodeless order parameter is preferable. Thus, to satisfy the Pauli principle, the order parameter must transform as a singlet under orbital rotations. This implies that the effective attraction favors pairing in the channels
$\Delta_{i,\sigma} = c_{i1\sigma} c_{i2\sigma}$ as well as
their linear superpositions. These are the pairing channels which are most favored by the ferromagnetic fluctuations.
As shown in Sec.~\ref{sec:superconductivity} of the main text, our QMC results show that indeed, pairing correlations in these channels are enhanced near the FM QCP.

\section{Possible superconducting phases}
\label{app:exotic_sc}
The two-component nature of the superconducting order parameter Eq.~\eqref{eq:SC_OP} may give rise to a number of exotic phases. Although our numerical data does not show evidence of such phases, in this section we describe how these phase could be detected.
We restrict our attention to the paramagnetic phase, since the two components, $\Delta_\uparrow$, $\Delta_\downarrow$, are related by the $Z_2$ symmetry of the model.
Define $\Delta_{\sigma}=\Delta e^{i\theta_{\sigma}}$, and switch to the charge/spin basis,
\begin{equation}
    \begin{split}
\theta_{\uparrow}&=\theta_{c}+\theta_{s}\\
\theta_{\downarrow}&=\theta_{c}-\theta_{s}.
\end{split}
\end{equation}
Neglecting amplitude fluctuations, the classical phase action is therefore $S=S_{c}+S_{s}$, with
\begin{equation}
    \label{eq:phase_action}
    S_{c,s}=\frac{1}{2}\int d^{2}rK_{c,s}(\nabla\theta_{c,s})^{2},
\end{equation}.
The possible phases (apart from a disordered phase) are:
\begin{enumerate}
\item Charge 4e superconductor: $\theta_{c}$ is quasi-long range ordered,
while $\theta_{s}$ is short ranged. An appropriate order parameter is $\Delta_{\uparrow}\Delta_{\downarrow}\propto e^{i2\theta_{c}}$,
whose correlations go as $\left\langle(\Delta_{\uparrow}^{\dagger}\Delta_{\downarrow}^{\dagger})(\mathbf{r})(\Delta_{\downarrow}\Delta_{\uparrow})(0)\right\rangle\propto r^{-\frac{4}{2\pi K_{c}}}.$
\item Spin-nematic: $\theta_{s}$ is quasi-long range ordered while $\theta_{c}$
is short ranged. The order parameter is $\Delta_{\uparrow}^{\dagger}\Delta_{\downarrow}\propto e^{-i2\theta_{s}}$,  whose correlations go as $\left\langle(\Delta_{\uparrow}^{\dagger}\Delta_{\downarrow})(\mathbf{r})(\Delta_{\downarrow}^{\dagger}\Delta_{\uparrow})(0)\right\rangle\propto r^{-\frac{4}{2\pi K_{s}}}.$
\item Triplet superconductor: both sectors have quasi-long range order. Here $\Delta_\uparrow$ has power law correlations, $\left\langle\Delta
^\dagger_\uparrow(\mathbf{r}) \Delta_\uparrow(0)\right\rangle \propto r^{-\left(\frac{1}{2\pi K_c}+\frac{1}{2\pi K_s}\right)}$.
\end{enumerate}

The phase diagram of the same model (in different physical contexts) has been studied in Refs.~\cite{Kruger2002,Podolsky2009,Berg2009}.
To determine the phase diagram, we consider the criteria for stability
against the appearance of a single vortex. In this model there are
three kinds of vortices:
\begin{enumerate}
\item A vortex of one spin species and an antivortex in the other. Across
the branch cut, $\theta_{c}\rightarrow\theta_{c}$, $\theta_{s}\rightarrow\theta_{s}+2\pi$.
\item A vortex of both the spin species. Across the branch cut, $\theta_{c}\rightarrow\theta_{c}+2\pi$,
$\theta_{s}\rightarrow\theta_{s}$.
\item A vortex of one of the spin species. Across the branch cut, $\theta_{c}\rightarrow\theta_{c}+\pi$,
$\theta_{s}\rightarrow\theta_{s}+\pi$.
\end{enumerate}
The phase diagram can be derived from considering the free energy of a single unpaired vortex, given by $F=E-TS$. (As in the usual Berezinski-Kosterlitz-Thouless transition, such an analysis reproduces the phase diagram from a more rigorous renormalization group treatment.) The stability condition
is $F>0$. The energy of a vortex of type 3 is
\[
E_{3}=\frac{1}{2}(K_{c}+K_{s})\int d^{2}r\left(\frac{\pi}{2\pi r}\right)^{2}=\frac{\pi}{4}(K_{c}+K_{s})\log(\frac{L}{a}),
\] where $L$ is the system size and $a$ is some short range cutoff.
Similarly $E_{2}=\pi K_{s}\log(\frac{L}{a})$, and $E_{1}=\pi K_{c}\log(\frac{L}{a})$. The entropy is the same in all cases, $TS=\log(\frac{L^{2}}{a^{2}})$.

\begin{figure}[th]
\begin{centering}
\includegraphics[width=0.7\columnwidth]{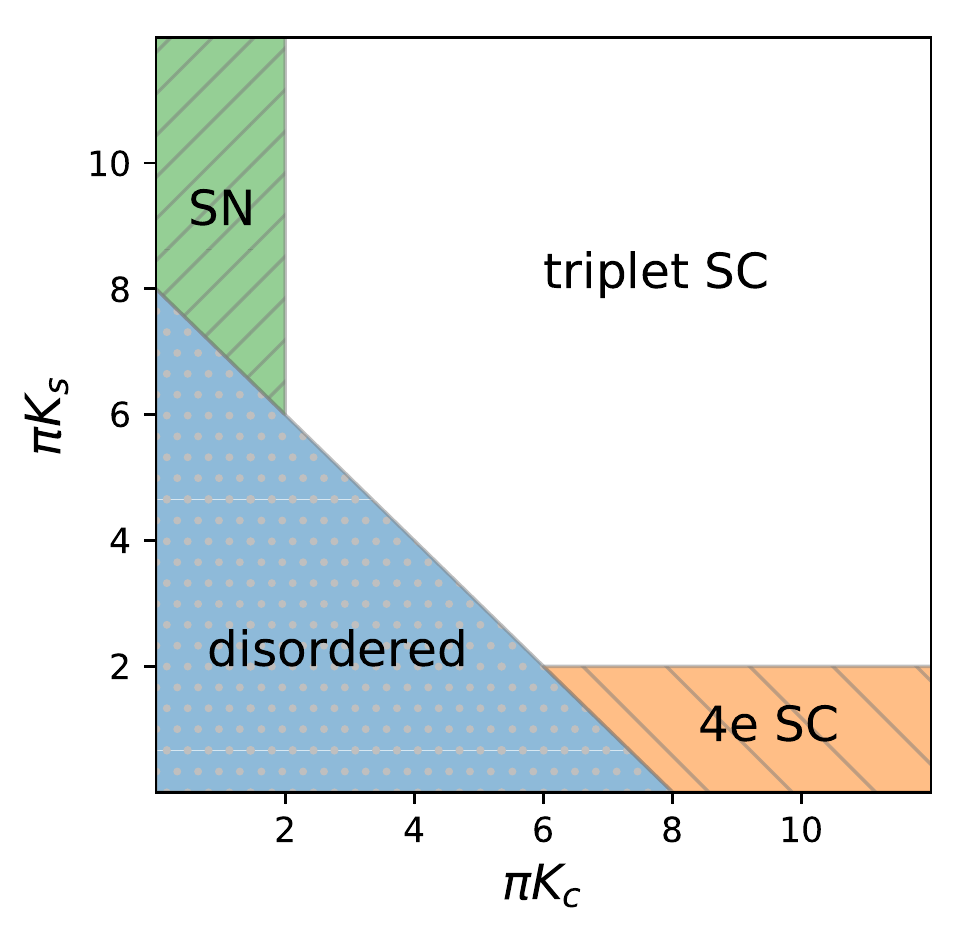}\caption{Phase diagram of the phase action~\eqref{eq:phase_action}, illustrating the spin-nematic (SN) phase, the charge $4e$ superconducting (4e SC) phase and the triplet superconducting phase.}\label{fig:sc_phase_diagram_theory}
\par\end{centering}
\end{figure}

The resulting phase diagram is given in Fig.~\ref{fig:sc_phase_diagram_theory}.

The stiffnesses $K_c$ and $K_s$ can be extracted from certain current-current correlation functions \cite{Scalapino1993},
\begin{equation}
\label{eq:stiffnesses}
    \begin{split}
    K_c&=\frac{\beta}{4}(\delta\Lambda_{\uparrow \uparrow} +\delta\Lambda_{\downarrow \downarrow} +2\delta\Lambda_{\uparrow \downarrow})\\
    K_s&=\frac{\beta}{4}(\delta\Lambda_{\uparrow \uparrow} +\delta\Lambda_{\downarrow \downarrow} -2\delta\Lambda_{\uparrow \downarrow}).
    \end{split}
\end{equation}

Here,
\begin{equation}
    \begin{split}
    \delta\Lambda_{\sigma,\sigma'} = \lim_{L\rightarrow \infty} \big[ &\Lambda^{xx}_{\sigma, \sigma'}(q_x=\frac{2\pi}{L}, q_y=0) - \\
    &\Lambda^{xx}_{\sigma, \sigma'}(q_x=0, q_y=\frac{2\pi}{L})\big],
    \end{split}
\end{equation}

where
\[
\Lambda^{xx}_{\sigma, \sigma'}(q_x, q_y)=\int d\tau \sum_{ij\lambda\lambda'}e^{i\mathbf{q}(\mathbf{r}_i-\mathbf{r}_j)} \left \langle j^x_{i\lambda \sigma}(\tau) j^x_{j \lambda' \sigma'} (0)\right\rangle,
\]
and $j^x_{i\lambda\sigma}=it e^{i\phi_{i,i+\hat{x}} ^{\lambda \sigma}} c^\dagger_{i\lambda\sigma}c_{i+\hat{x},\lambda\sigma} + \mathrm{H.c}$ is the current density for fermions of orbital $\lambda$ and spin $\sigma$.

\section{Superfluid density and pairing correlations}
\label{app:superfluiddensity}
In this appendix we provide further details on the numerical evidence for superconductivity.

\begin{figure}
\includegraphics[width=\columnwidth]{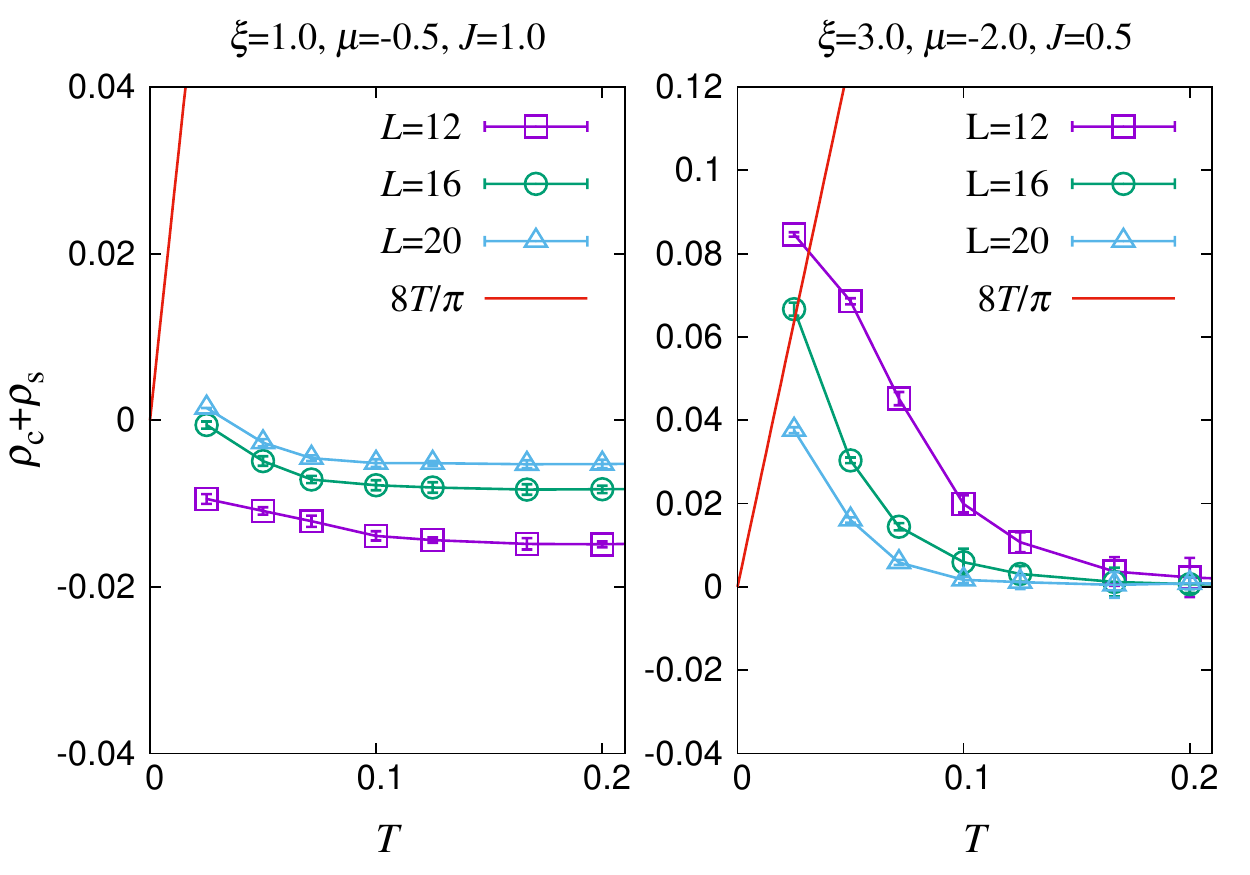}
\caption{Temperature dependence of $\rho_c+\rho_s$ at QCP for different system sizes and different couplings.}
\label{fig:Kt}
\end{figure}

\begin{figure}
\includegraphics[width=\columnwidth]{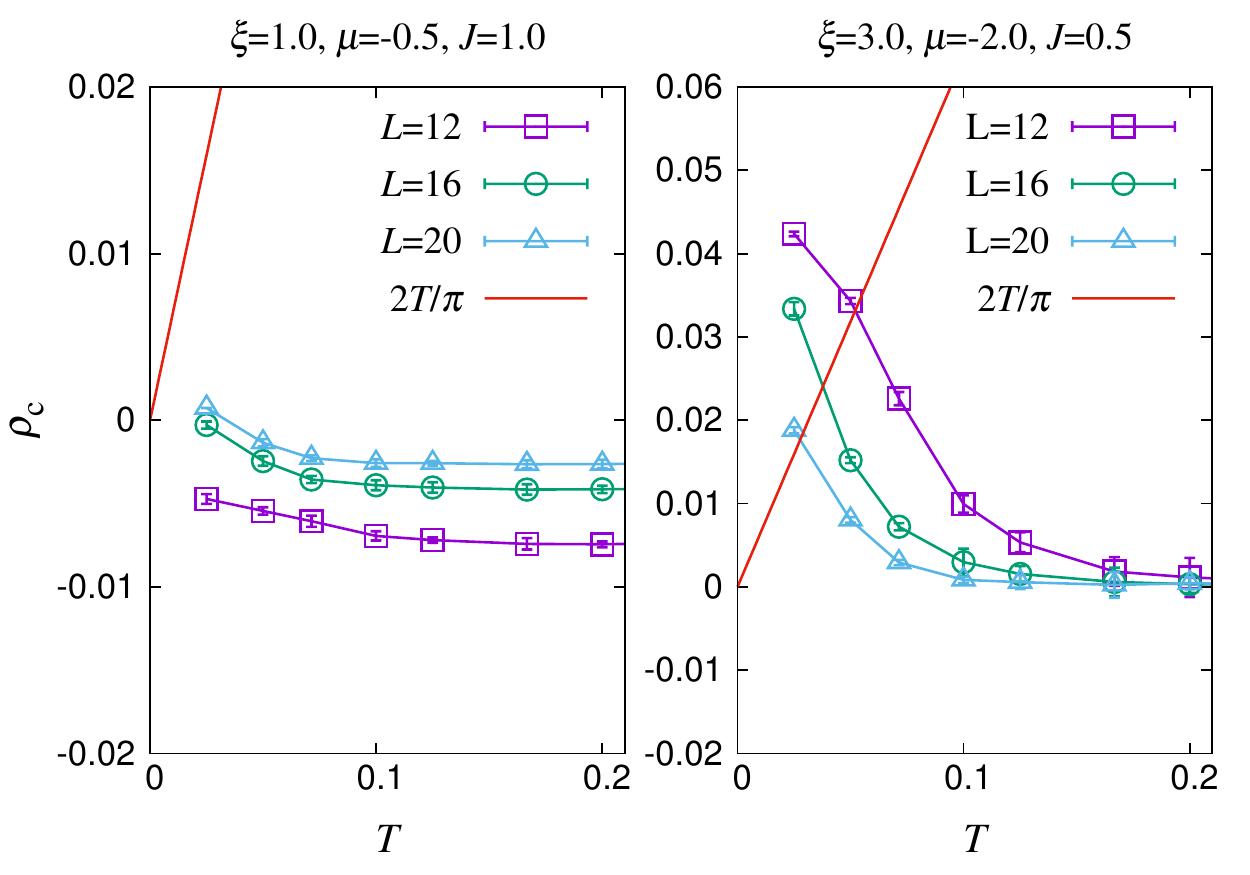}
\caption{Temperature dependence of $\rho_c$ at QCP for different system sizes and different couplings.}
\label{fig:Kc}
\end{figure}

\begin{figure}
\includegraphics[width=\columnwidth]{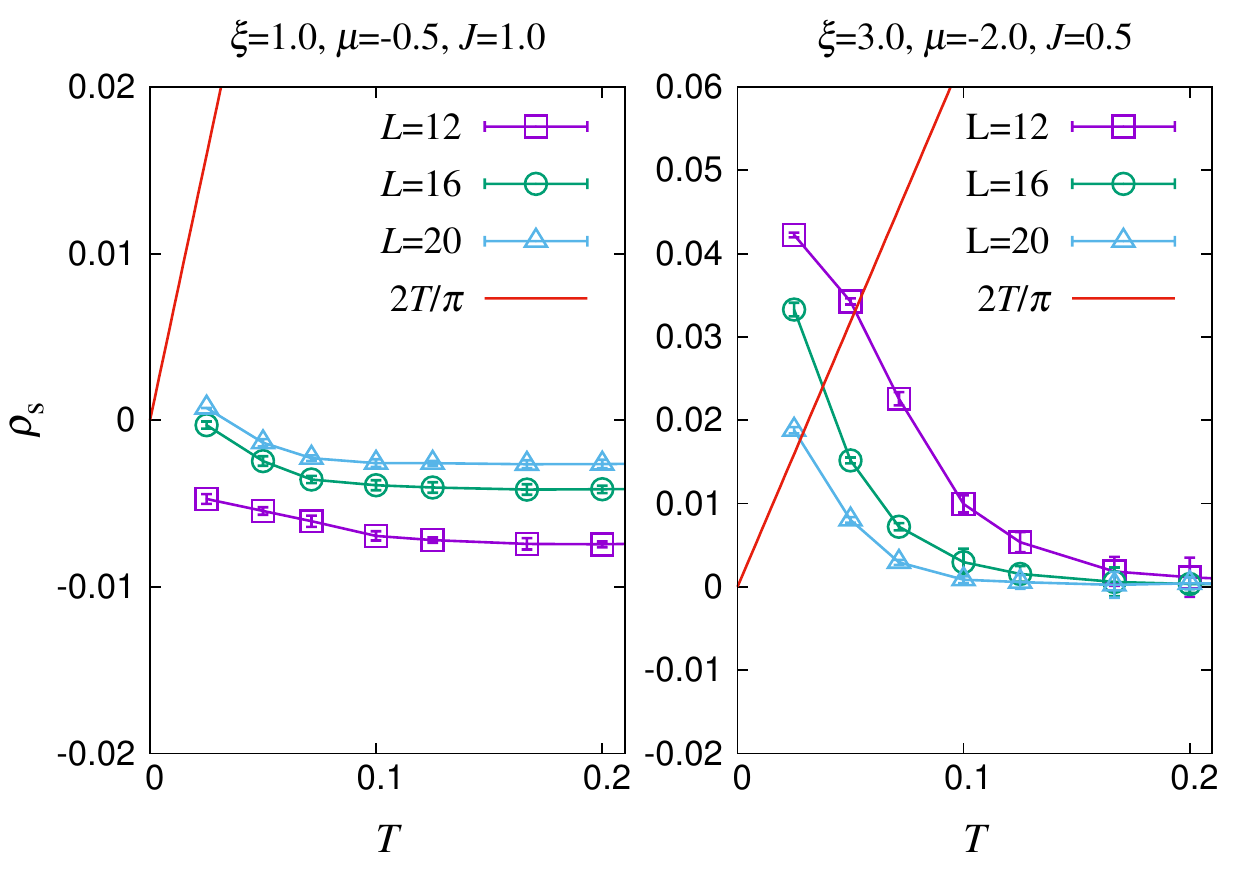}
\caption{Temperature dependence of $\rho_s$ at QCP for different system sizes and different couplings.}
\label{fig:Ks}
\end{figure}

To identify the leading pairing channel, we considered all possible on-site and nearest neighbor pairing order parameters and computed the pair structure factor for each channel. In all channels other than the orbital-singlet, spin-triplet channel defined in Eq.~\eqref{eq:SC_OP} of the main text, we find a weak response with no substantial system-size dependence or enhancement close to the FM QCP (not shown).

To test whether there are possible superconductivity instabilities close to the QCP, we measured the superfluid densities $\rho_c$ and $\rho_s$ which are related to the stiffnesses $K_c$ and $K_s$ defined in Eq.~\eqref{eq:stiffnesses}. $\rho_{c,s}=K_{c,s}/\beta$.
In Fig. ~\ref{fig:Kt} we show the temperature dependence of $\rho_c+\rho_s$. For both sets of parameters, $\rho_c+\rho_s <\frac{8}{\pi} T$ at the largest system size, and it decreases with the system size, implying the absence of quasi-long range superconducting order down to $T=0.025$ (see Fig.~\ref{fig:sc_phase_diagram_theory}).
Note that upon increasing coupling strength and decreasing temperature, the finite-size estimates for the superfluid density grow. It is likely that there is a transition to a superconducting phase at higher coupling strengths or lower temperatures.

For completeness, in Fig.~\ref{fig:Kc}, ~\ref{fig:Ks} we show $\rho_c$, $\rho_s$, respectively.

\section{Fermion spin susceptibility}
\label{app:fermionsusceptibility}

\begin{figure}
\includegraphics[width=\columnwidth]{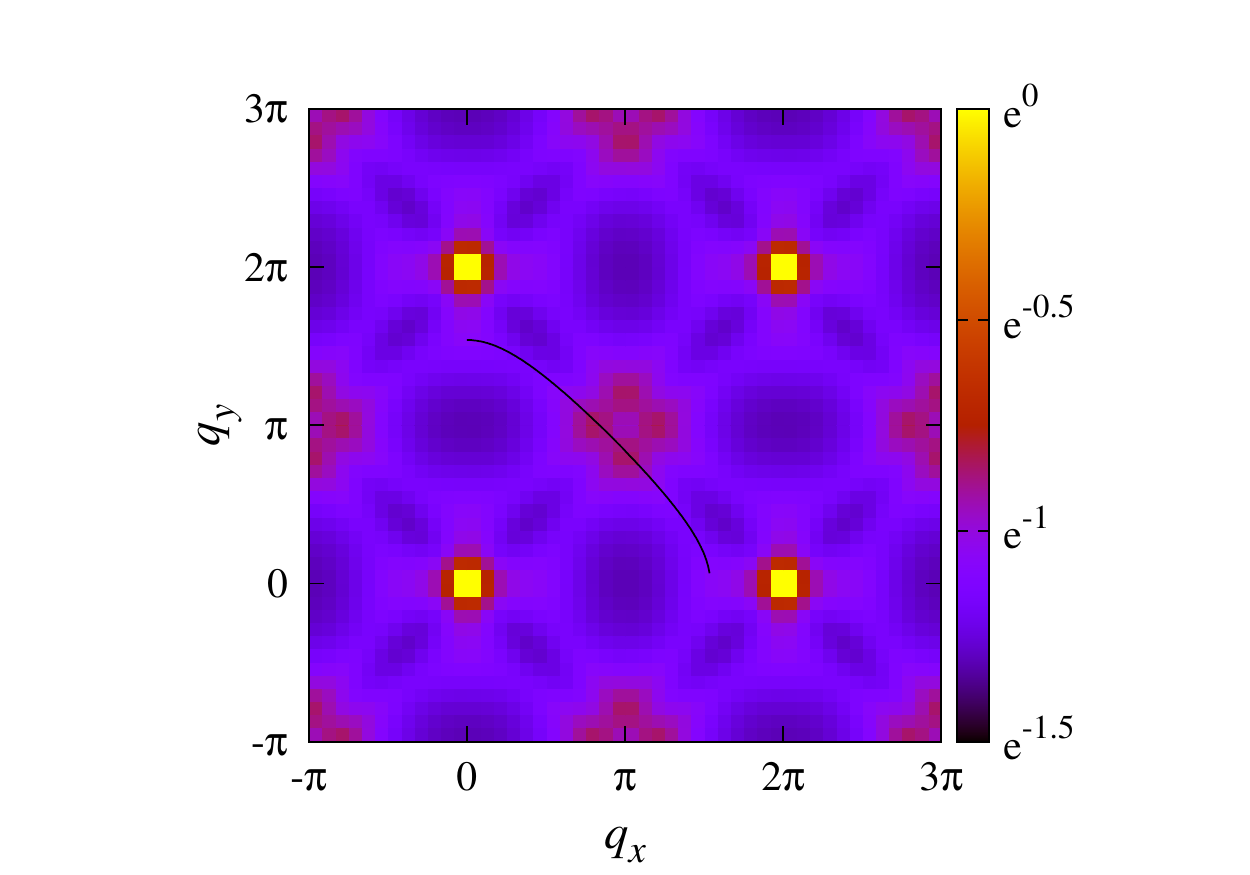}
\caption{ The intensity distribution of fermion spin susceptibility  $\chi_\mathrm{F} (h=h_c,T=0.05,\vec{q},\omega_n=0)$ with $L=24$.  To amplify the weak signal of the soft mode, we have fixed the intensity range less than unity and taken a logarithm of original data. The black line shows the location of $2\vec{k}_F$ for the noninteracting FS,  only a quarter of which  are plotted here.}
\label{fig:chi_qmap}
\end{figure}

In the Fig.~\ref{fig:zt_sig}(a) of the main text, we discussed the anisotropy of the quasiparticle fraction $Z_{\mathbf{k}_{F}}(T)$ along $\theta=0$ and $\theta=\frac{\pi}{4}$ directions, and associate such anisotropy to the soft fermion-bilinear mode around $\mathbf{Q}=(\pi,\pi)$ in the fermion spin susceptibility, besides the dominant $\mathbf{Q}=(0,0)$ ferromagnetic fluctuations. Here we show the fermion spin susceptibility (Eq.\eqref{eq:fermionspinsuscep}) in Fig.~\ref{fig:chi_qmap}, and reveal it is indeed the case.

Fig.~\ref{fig:chi_qmap} shows the fermion spin susceptibility at $h=h_c$, $T=0.05$ from a $L=24$ system. The strongest intensity is naturally at $\mathbf{Q}=(0,0)$, however, near $\mathbf{Q}=(\pi,\pi)$ , there are soft modes (the signal is very weak and we have to fix the intensity range less than one and take a logarithm of original data). %We indeed see some anisotropy of the soft modes which come from the anisotropy of FS and explains the anisotropy of quasi-particle weight we observed at QCP.

\section{Self energy}
\label{app:self_energy}
As shown in Fig.~\ref{fig:zt_sig} in the main text, the imaginary part of the fermionic self-energy $-\mathrm{Im}\Sigma(\omega_n)$ increases in magnitude as the frequncy is lowered. While such behavior is expected in a superconducting state, the superconducting fluctuations were found to be extremely weak. Here we speculate about the possible mechanism for such behavior.
%EBpropose one possible mechanism for such behavior, but stress that at present, its relevance to the model at hand is highly speculative.

Close to the classical Ising transition, the dynamics of the order parameter are slow. We focus on the classical fluctuations of the Ising spins, i.e their static configurations. For each such static configuration, one may solve the fermionic part of the Hamiltonian, and obtain the single-particle Green's function. Let us further assume that the fluctuations are extremely sharply peaked at $q=0$. Then, in a Monte-Carlo simulation, the configurations alternate between static, nearly spatially uniform configurations of Ising spins. Most importantly, the sign of the order parameter changes between these configurations, or else the system is by definition in the ordered phase. Hence, neglecting all other interaction effects, the fermionic Green's function takes the form
\begin{equation}
    G(\mathbf{k}, {i\omega_n}) \approx \frac{1}{2} \left(\frac{1}{i\omega_n - \epsilon_{k} -\Delta} + \frac{1}{i\omega_n - \epsilon_{k} + \Delta} \right),
\end{equation}
where $\Delta$ is the magnitude of the local ferromagnetic order parameter. If $\epsilon_k\ll \Delta,\omega$, at the nominal Fermi surface ($\epsilon_k=0$), we find that
\[
-\mathrm{Im} \Sigma(i\omega_n) = \mathrm{Im}\left[G^{-1}(\mathbf{k},i\omega_n)\right] -\omega_n = \frac{\Delta^2}{\omega_n},
\]
which increases rapidly as the frequency is lowered.
If $\Delta$ were large, compared, e.g to temperature, it would be possible to measure it by other means. This would also imply $|\mathrm{Im} \Sigma(i\omega_0)|\gg \omega_0=\pi T$.  Small gaps, however, are difficult to detect.

\bibliography{main}

\end{document}